\documentclass[twocolumn]{aastex631}

\usepackage[croatian,english]{babel}

\def \grb {\mbox{GRB\,221009A}}
\def \xmm {\emph{XMM-Newton}}

\def \ixpe {\emph{IXPE}}
\def \fermi {\emph{Fermi}}
\def \sw  {\emph{Swift}}
\def \swi {\emph{Swift}}

\begin{document}

\title{The power of the rings: the GRB 221009A soft X-ray emission from its dust-scattering halo}

\author[0000-0002-6038-1090]{Andrea Tiengo}
\affiliation{Scuola Universitaria Superiore IUSS Pavia, Piazza della Vittoria 15, 27100 Pavia, Italy}
\affiliation{Istituto Nazionale di Astrofisica, Istituto di Astrofisica Spaziale e Fisica Cosmica di Milano, via A. Corti 12, 20133 Milano, Italy}
 
\author[0000-0002-3869-2925]{Fabio Pintore}
\affiliation{Istituto Nazionale di Astrofisica, Istituto di Astrofisica Spaziale e Fisica Cosmica di Palermo, Via U. La Malfa 153, 90146 Palermo, Italy}
 
\author[0000-0003-0852-0257]{Beatrice Vaia}
\affiliation{Scuola Universitaria Superiore IUSS Pavia, Piazza della Vittoria 15, 27100 Pavia, Italy}
\affiliation{
Department of Physics, University of Trento, Via Sommarive 14, 38123 Povo (TN), Italy}
\affiliation{Istituto Nazionale di Astrofisica, Istituto di Astrofisica Spaziale e Fisica Cosmica di Milano, via A. Corti 12, 20133 Milano, Italy}
  
\author[0000-0002-2119-9835]{Simone Filippi}
\affiliation{Scuola Universitaria Superiore IUSS Pavia, Piazza della Vittoria 15, 27100 Pavia, Italy}
\affiliation{Department of Physics “A.\,Volta”, University of Pavia, Via Bassi 6, 27100 Pavia, Italy}
  
\author[0000-0002-7295-5661]{Andrea Sacchi}
\affiliation{Scuola Universitaria Superiore IUSS Pavia, Piazza della Vittoria 15, 27100 Pavia, Italy}

\author[0000-0003-4849-5092]{Paolo Esposito}
\affiliation{Scuola Universitaria Superiore IUSS Pavia, Piazza della Vittoria 15, 27100 Pavia, Italy}
\affiliation{Istituto Nazionale di Astrofisica, Istituto di Astrofisica Spaziale e Fisica Cosmica di Milano, via A. Corti 12, 20133 Milano, Italy}

\author[0000-0001-6641-5450]{Michela Rigoselli}
\affiliation{Istituto Nazionale di Astrofisica, Istituto di Astrofisica Spaziale e Fisica Cosmica di Milano, via A. Corti 12, 20133 Milano, Italy}

\author[0000-0003-3259-7801]{Sandro Mereghetti}
\affiliation{Istituto Nazionale di Astrofisica, Istituto di Astrofisica Spaziale e Fisica Cosmica di Milano, via A. Corti 12, 20133 Milano, Italy}

\author[0000-0002-9393-8078]{Ruben Salvaterra}
\affiliation{Istituto Nazionale di Astrofisica, Istituto di Astrofisica Spaziale e Fisica Cosmica di Milano, via A. Corti 12, 20133 Milano, Italy}

\author[0000-0001-5391-8286]{Barbara \v Siljeg}
\affiliation{ASTRON, the Netherlands Institute for Radio Astronomy, Oude Hoogeveensedijk 4, 7991 PD Dwingeloo, The Netherlands}
\affiliation{Kapteyn Astronomical Institute, University of Groningen, P.O. Box 800, 9700 AV, Groningen, The Netherlands}
\affiliation{Ru{\dj}er Bo\v{s}kovi\'c Institute, Bijeni\v{c}ka cesta 54, 10000 Zagreb, Croatia}

\author[0000-0003-0932-3140]{Andrea Bracco}
\affiliation{Laboratoire de Physique de l'Ecole Normale Sup\'erieure, ENS, Universit\'e PSL, CNRS, Sorbonne Universit\'e, Universit\'e de Paris, 75005 Paris, France}

\author[0000-0001-6536-0320]{\v Zeljka Bo\v snjak}
\affiliation{Faculty of Electrical Engineering and Computing, University of Zagreb, Unska ul.3, 10000 Zagreb, Croatia}

\author[0000-0002-6034-8610]{Vibor Jeli\' c}
\affiliation{Ru{\dj}er Bo\v{s}kovi\'c Institute, Bijeni\v{c}ka cesta 54, 10000 Zagreb, Croatia}

\author[0000-0001-6278-1576]{Sergio Campana}
\affiliation{Istituto Nazionale di Astrofisica, Osservatorio Astronomico di Brera, Via E. Bianchi 46, 23807 Merate (LC), Italy}







\begin{abstract}
\grb\ is the brightest gamma-ray burst (GRB) ever detected and occurred at low Galactic latitude.
Owing to this exceptional combination, its prompt X-ray emission could be detected for weeks in the form of expanding X-ray rings produced by scattering in Galactic dust clouds. We report on the analysis of 20 rings, generated by dust at distances ranging from 0.3 to 18.6 kpc, detected during two 
\xmm\ observations performed about 2 and 5 days after the GRB.
By fitting the spectra of the rings with different models for the dust composition and grain size distribution, we reconstructed the spectrum of the GRB prompt emission in the 0.7--4 keV energy range as an absorbed power law with photon index $\Gamma=1$--1.4 and absorption in the host galaxy $N_{\rm H,z}=(4.1$--$5.3)\times10^{21}$\,cm$^{-2}$. Taking into account the systematic uncertainties on the column density of dust contained in the clouds producing the rings, the  0.5--5 keV  fluence of \grb\ can be constrained between 10$^{-3}$ and $7\times10^{-3}$\,erg\,cm$^{-2}$.
\end{abstract}
\keywords{Gamma-ray bursts(629) --- Interstellar dust(836) --- Interstellar dust extinction(837) --- Interstellar scattering(854)}

\section{Introduction}\label{sec:intro}
The gamma-ray burst (GRB) 221009A was first reported when its  hard X-ray ($>$15 keV) emission triggered the \swi\,/BAT instrument on 2022 October 9 at 14:10:17 UT  \citep{2022GCN.32632....1D}, but, owing to its low Galactic latitude and long duration, it was initially believed to be a new Galactic transient. However, it was soon realized that it was the afterglow   of a very bright and long GRB, which had been detected  with the Gamma-Ray Burst Monitor (GBM, 10\,keV--25\,MeV) onboard \fermi\ about one hour earlier \citep{2022GCN.32636....1V,2022GCN.32635....1K}.\footnote{\swi\ was occulted by the Earth at the time of the GRB onset.}

\grb\ is  the  brightest GRB ever observed \citep{williams23}: most of the X-/$\gamma$-ray instruments were saturated (or strongly affected by dead time or pile-up) 
during  the peak of its emission. Such an exceptional brightness  resulted from the combination of a rather high  (but not extreme) intrinsic luminosity and a  close  distance (745 Mpc; \citealt{malesani23}). 
It has been estimated that the occurrence of a burst of this luminosity within this distance 
might be once every 1,000 years \citep{williams23,malesani23}.

\grb\ occurred close to the direction of the Galactic Plane ($b$=4$\fdg32$ and $l$=52$\fdg96$) and, remarkably, produced several bright X-ray scattering rings caused by interstellar dust in our Galaxy \citep{2022GCN.32680....1T,williams23}. Dust scattering rings  have been observed so far only in a handful of GRBs \citep{vaughan04,tm06,vaughan06,vianello07, pintore17grb}. Their study is useful to derive information on the dust (e.g. its distance, composition and grain size distribution) as well as on the properties of the GRB prompt emission in the soft X-ray range, which usually cannot be observed directly. 

Here we report on the dust scattering rings of \grb\ as observed with the \xmm\ satellite. 
Our  main objective is to estimate fluence and spectral shape of the prompt emission in the soft X-ray range (0.7--4 keV). A more complete exploitation of this exceptional data set to characterize the dust properties is deferred to forthcoming publications.

The distances of the scattering dust layers can be derived from the angular expansion rate of the rings, caused by the longer path length of the  photons scattered at larger angles  \citep{truemper73}. On the other hand,  inferring the GRB properties from the observed rings is more complicated and requires some assumptions. Indeed, since the specific intensity of the scattered radiation is proportional to the product of the burst fluence and dust optical depth, some independent information is required to break the degeneracy between these two quantities. Thanks to the high statistical quality of our X-ray observations, which allow us to measure the spectra of a large number of rings as well as their time evolution and azimuthal brightness variations, we could compare the results obtained with various methods and assumptions, thus deriving the spectrum of the GRB prompt emission in the soft X-ray range with a realistic estimate of the systematic uncertainties.
In particular, we considered the cross sections corresponding to eight different  models of interstellar dust grains \citep{draine03,zubko04,mathis77}. We also used various studies of optical extinction in this direction: three-dimensional (3D) maps   to estimate the contributions of individual dust layers \citep{green19,lallement22} and two-dimensional (2D) maps  to derive the integrated absorption along the whole line of sight \citep{schlafly11,planck14}.

\section{Observations}

The first \xmm\ observation (Obs.ID 0913991501; hereafter Obs1) started on October 11, 2022 at 21:11:10, about 2.3 days after the GRB trigger and lasted about 50 ks. The main aim of this observation was the study of the X-ray afterglow, which was still extremely bright. Therefore, to minimise the photon pile-up, the EPIC PN \citep{struder01} and the two EPIC MOS \citep{turner01} cameras were operated in Timing mode, with the Thick optical-blocking filter. 
In Timing mode, full imaging capabilities are provided only by the peripheral CCDs of the MOS cameras.\footnote{Six CCDs in MOS2 and only four in MOS1, because two of them were permanently damaged in the early phases of the mission.}
Since the size of each MOS CCD is 11$^{\prime}\times11^{\prime}$, X-ray rings with radii smaller than $5\farcm5$
could not be observed, and only MOS2 could image the whole azimuthal extent of the rings.
The particle background remained low and constant throughout the whole observation, resulting in a net exposure time of 47.8 ks for the MOS2. 

A second observation (Obs.ID 0913991601; hereafter Obs2), with a total duration of 62 ks, was performed on October 14, 2022. Since it was mainly devoted to the study of the rings, all the EPIC cameras were operated in Full Frame mode, which provides full imaging capabilities over the whole field of view ($\sim$14$^{\prime}$ radius), and the  Thin optical  filter was used. The second part of this observation was affected by strong and variable particle background. Therefore, we limited the analysis to the longest uninterrupted time interval with quiescent background for both the PN and MOS2 cameras.\footnote{MOS1 data were also checked for consistency, but MOS1 results are not reported because the two missing CCDs did not allow us to fully cover the rings with a radius larger than $5\farcm5$.} 
This resulted in a net exposure time of 29.7 ks for the PN and 33.5 ks for the MOS2.

Further \xmm\     observations, carried out 21, 23 and 32 days after the GRB, are crucial to    better characterize the properties of the dust clouds located on the other side of the Galaxy,
but we will not use them because, due to the faintness of the rings, they do not give a substantial contribution to the reconstruction of the GRB prompt X-ray emission, which is the main aim of the present work. 

\begin{figure*}
    \centering
    \includegraphics[width=.49\textwidth]{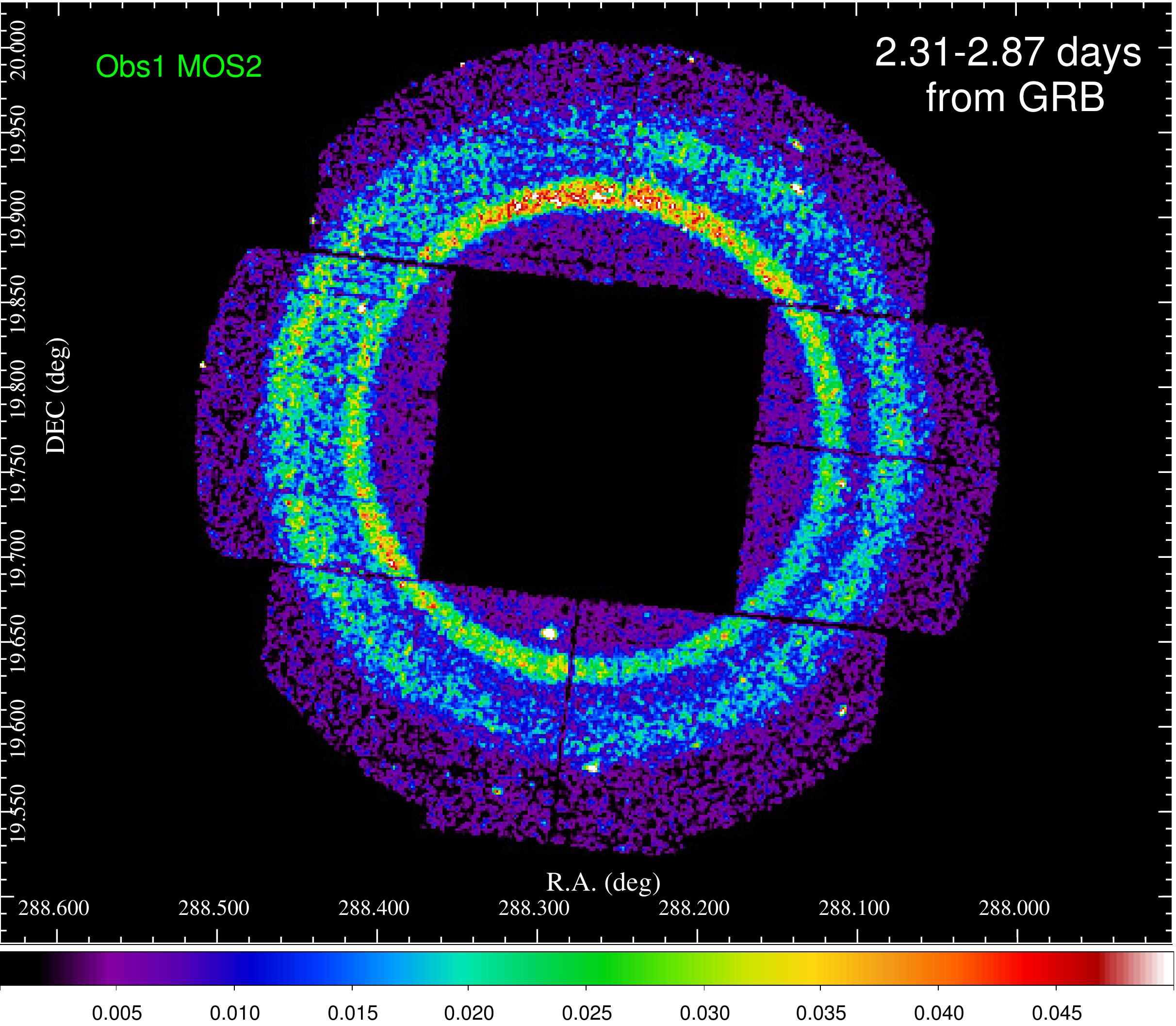}
    \includegraphics[width=.49\textwidth]{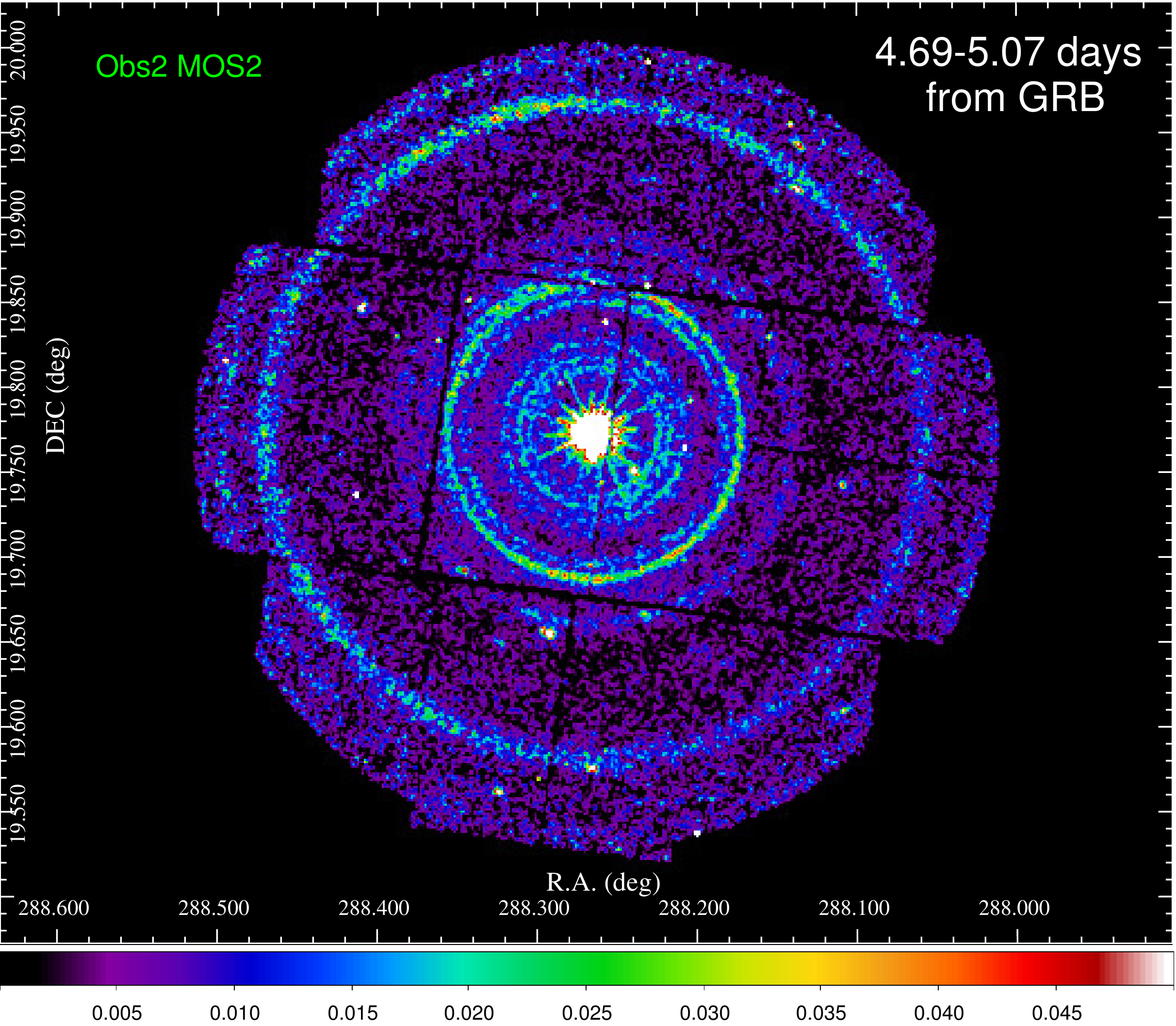}
\caption{EPIC-MOS2 exposure-corrected 0.7--4 keV images, in units of counts s$^{-1}$ arcmin$^{-2}$, of the expanding rings from Obs1 (left panel) and Obs2 (right panel). All the images have been smoothed with a Gaussian kernel of $\sigma=3\farcs5$.}
    \label{fig:ringsmos2}
\end{figure*}

\section{Data analysis and Results} 

The  data were processed using \texttt{SAS 19.1.0} \citep{gabriel04} and the most recent calibration files. 
The EPIC events were cleaned using standard filtering expressions.
To maximise the signal-to-noise ratio of the rings, the analysis was restricted to the 0.7--4 keV energy band.

The MOS2 images of the two observations are shown in Fig.\,\ref{fig:ringsmos2}.
Some of the X-ray rings that were visible in the first observation (left panel) have moved (partly) outside the field of view in the second one (right panel), but several additional rings, produced by dust clouds at larger distances, have become accessible 
at radii $<$6$^{\prime}$. 
These  images show that the surface brightness of the individual rings is not uniform (see also Appendix~\ref{images}), which indicates  that the  dust clouds responsible for the scattering are not homogeneous in the plane of the sky. 

A number of point-like sources are apparent and for the ring analysis
they were removed by excluding   circular regions of radius $\sim$$20''$. Out-of-time events were also removed following the standard procedure.\footnote{See \url{https://www.cosmos.esa.int/web/xmm-newton/sas-thread-epic-oot}.}

\subsection{Galactic dust distribution}\label{PDsec}

We studied the expanding dust-scattering rings by adopting the approach of the so-called {\it pseudo-distance} distribution \citep{tm06}.
For each detected event we compute the  quantities
\begin{equation}
t_{\rm i} = T_{\rm i} - T_{\rm 0} 
\end{equation}
\begin{equation}
\theta_{\rm i}^2 = (x_{\rm i} - X_\mathrm{GRB})^2 + (y_{\rm i} - Y_\mathrm{GRB})^2 
\end{equation}
where $x_{\rm i}$, $y_{\rm i}$, and $T_{\rm i}$ are the detector coordinates and the time of arrival of the i-event, $T_{\rm 0}$, $X_\mathrm{GRB}$ and $Y_\mathrm{GRB}$ are the GRB start time and detector coordinates. $T_{\rm 0}$ is fixed at MJD=59861.55568,
when the bulk of the prompt emission of \grb\ was detected, and the GRB coordinates are derived by taking the central position of the afterglow in each observation. 
The {\it pseudo-distance} is  defined as: 

\begin{equation}
D_{\rm i} = 2ct_{\rm i} / \theta_{\rm i}^2 = (19.85 t_{\rm i} [\text{days}]) / (\theta_{\rm i}^2 [\text{arcmin}]) \text{~kpc}.
\label{PDformula}
\end{equation} 
\noindent For background events, this quantity is not a real distance.  
Instead, photons from an expanding ring have values of $D_{\rm i}$ that cluster around the distance of the scattering dust layer. 

\begin{figure}[ht!]
\includegraphics[width=9cm]{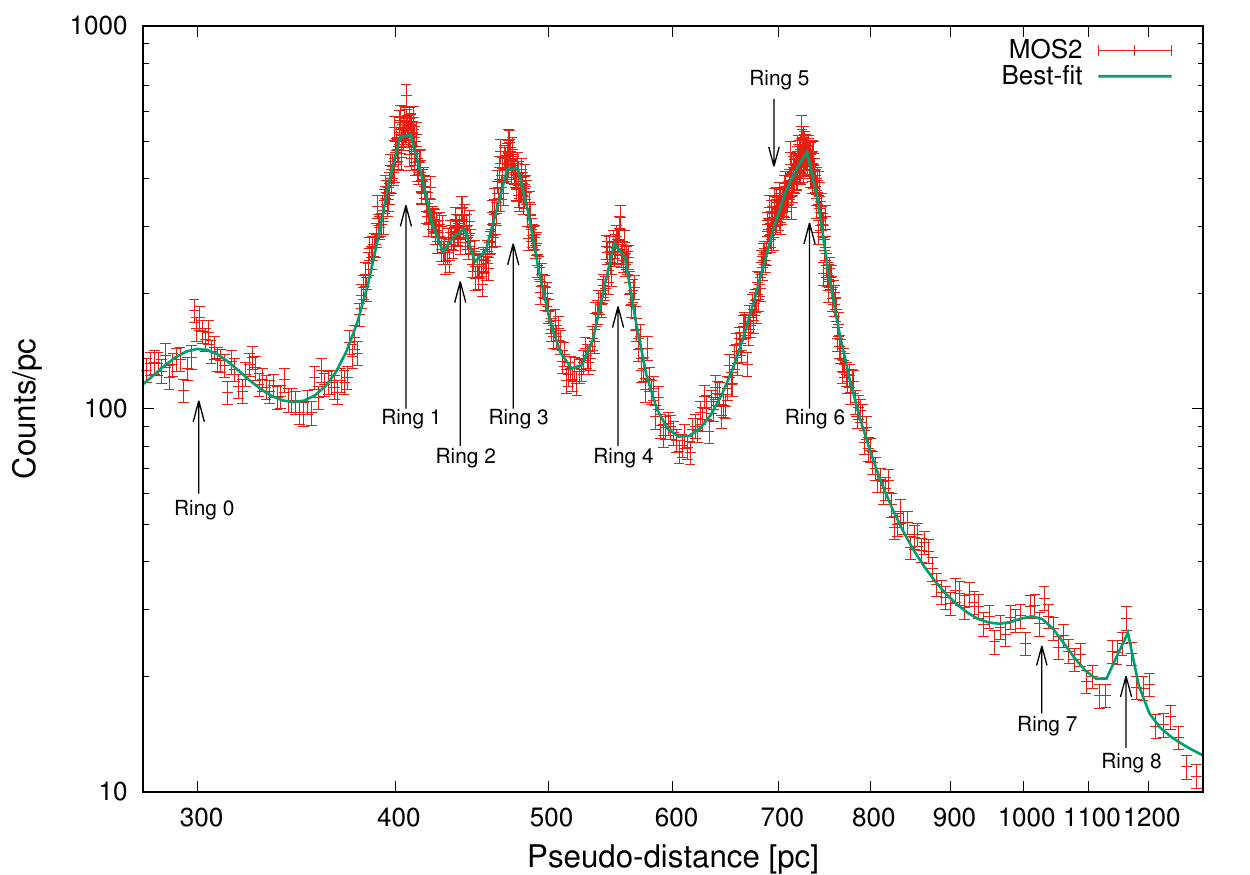}
\includegraphics[width=9cm]{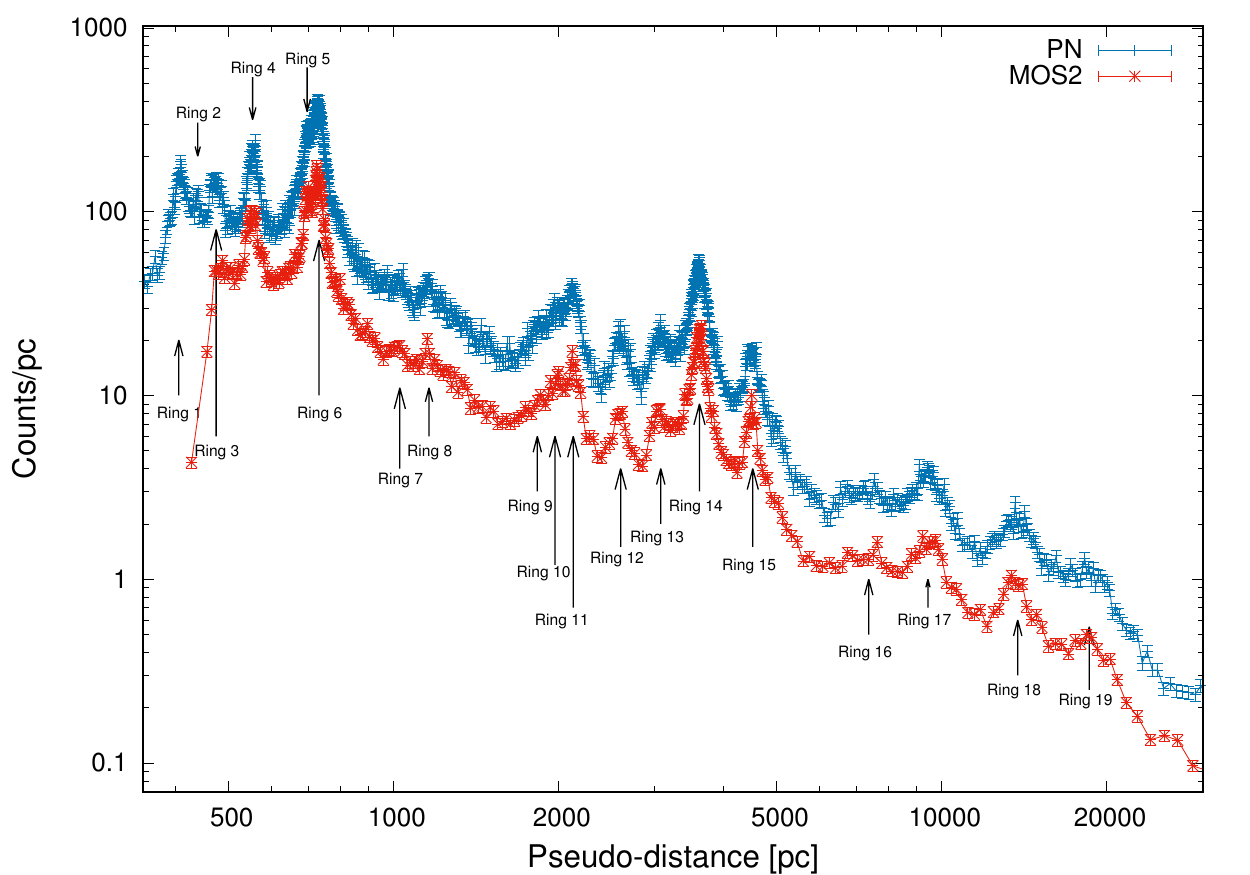}
\caption{{\it Pseudo-distance} distributions in the 0.7--4 keV energy band for the two \xmm\ observations ({\it Obs1: top;    Obs2: bottom}). Arrows indicate the number of the rings as reported in Table~\ref{DistTab}. MOS2 data are in red and PN in blue. The green line of the top panel indicates the best-fit model 
with a power law plus 9 Lorentzian functions. 
\label{PDs}}
\end{figure}

The histograms of {\it pseudo-distances} obtained in the two observations (Fig.\,\ref{PDs}) clearly show  a number of peaks that indicate the presence of many dust scattering layers at different distances.
We fit separately the histogram of each instrument with the sum of a power law (to account for the background) and a series of Lorentzian functions (9 for Obs1 and 19 for Obs2). 
The best-fit  values obtained for the peak centroids and widths\footnote{The width of a peak in the {\it pseudo-distance} distribution depends on the GRB duration (negligible in this case), the instrumental point-spread function (PSF, $\sim$20$^{\prime\prime}$ half-energy width), and the intrinsic thickness of the dust layer \citep{vianello07}.} are reported in Table~\ref{DistTab},
together with the angular radii ($\theta_1$ and $\theta_2$) of the rings at the beginning and at the end of each observation, derived from Eq.~\ref{PDformula}.

In total, we identified 20 dust layers, with the most-prominent ones clustered at distances in the range 400--750\,pc. We also located a group of dust layers between 2 and 5\,kpc and another one between 10 and 20\,kpc. 
Given the  Galactic latitude $b= 4\fdg3$, this implies the presence of dust  up to $\sim$1.5\,kpc above the Galactic Plane, in a region where it is difficult to obtain information on the interstellar dust studying the optical extinction.

\begin{table}
\label{ring_dist}
\centering                          
\begin{tabular}{lccc}
\hline\hline    
Ring & Distance  & Width  & $\theta_1-\theta_2$ in Obs1; Obs2\\   
 & [pc]  & [pc] & [arcmin] \\   
\hline   
0 & $300\pm2$  & $62\pm10$ & (12.41--13.81); (17.66--18.37)\\
1 & $406.3\pm0.2$  & $26.9\pm0.7$ & 10.67--11.86; (15.17--15.79)\\
2 & $439.8\pm0.5$  & $14.6\pm1.9$ & 10.25--11.40; (14.59--15.17)\\
3 & $475.2\pm0.3$  & $30.9\pm0.9$ & 9.86--10.97; (14.03--14.60)\\
4 & $553.6\pm0.3$  & $27.7\pm1.0$ & 9.14--10.16; (13.00--13.52)\\
5 & $695.4\pm1.2$  & $23.1\pm3.7$ & 8.15--9.07; 11.60--12.07\\
6 & $728.6\pm1.1$  & $42.7\pm2.5$ & 7.96--8.86; 11.33--11.79\\
7 & $1027.3\pm5.2$  & $38.1\pm8.7$ & (6.71--7.46); 9.54--9.93\\
8 & $1161.7\pm2.5$  & $99\pm21$ & (6.31--7.02); 8.97--9.34\\
9 & $1831\pm13$  & $121\pm44$ & (5.02--5.59); 7.15--7.44\\
10 & $1973\pm10$  & $151\pm52$ & (4.84--5.38); 6.89--7.16\\
11 & $2129\pm5$  & $135\pm14$ & (4.66--5.18); 6.63--6.90\\
12 & $2599\pm5$  & $164\pm18$ & (4.22--4.69); 6.00--6.24\\
13 & $3075.5\pm7.4$  & $309\pm28$ & (3.88--4.31); 5.52--5.74\\
14 & $3616.5\pm2.8$  & $308\pm9$ & (3.58--3.98); 5.09--5.29\\
15 & $4526.3\pm7.5$  & $494\pm28$ & (3.20--3.55); 4.55--4.73\\
16 & $7374\pm62$  & $1410\pm225$ & (2.50--2.79); 3.56--3.71\\
17 & $9465\pm32$  & $1584\pm108$ & (2.21--2.46); 3.14--3.27\\
18 & $13765\pm53$  & $2608\pm196$ & (1.83--2.04); 2.61--2.71\\
19 & $18592\pm128$  & $4712\pm442$ & (1.58--1.75); 2.24--2.33\\

\hline                                  
\end{tabular}\\
\caption{Ring distances, widths (FWHM of the Lorentzian peaks in Fig.\,\ref{PDs}) and expected radii in the two \xmm\ observations. Distances and widths from MOS2 (Obs1) for rings 0--4 and 7--8, from MOS2 (Obs2) for rings 5--6 , from PN (Obs2) for rings 9--19. The minimum ($\theta_1$) and maximum ($\theta_2$) radii expected during each exposure are indicated in parentheses   if no EPIC camera could fully image the ring and so its spectrum was not considered in the spectral fitting.
}
\label{DistTab}
\end{table}

\subsection{Spectral analysis and results}

The ring spectra were extracted by fitting  Lorentzian functions to the peaks in the energy-resolved {\it pseudo-distance} distributions, assuming a power-law model for the background contribution (as explained in \citealt{tm06} and \citealt{pintore17grb}). The Lorentzian centroids and widths were kept fixed to the values obtained, for each observation and instrument, in the whole 0.7--4 keV energy band  (Fig.\,\ref{PDs}). Their normalizations were instead derived in suitable energy intervals, chosen to have at least 100 net counts in each peak.
For each spectrum, response matrices were produced with the \texttt{SAS} tasks \texttt{rmfgen} and \texttt{arfgen} for the annular regions covered by each ring during the observation. The matrices are not corrected for the instrumental PSF, since the integration of the Lorentzian functions from which the spectral bins are generated already comprises all the background-subtracted counts in the ring.

The spectrum $F(E)$ of a ring produced by the single-scattering of GRB photons with energy $E$ in a geometrically thin dust cloud in our Galaxy, as observed to expand from $\theta_1$ to $\theta_2$, can be written as:
\begin{equation}
F(E)=\frac{f(E)}{T_{\rm exp}}\Delta N_{\rm H}\sigma_{\theta_{1,2}}(E),
\label{specmod}
\end{equation}

\noindent where $f(E)$ is the GRB specific fluence, $T_{\rm exp}$ is the observation duration, $\Delta N_{\rm H}$ is the equivalent hydrogen column density of dust in the cloud, and $\sigma_{\theta_{1,2}}(E)$ is the scattering cross-section between the angles $\theta_1$ and $\theta_2$.
The latter can be computed using the model by \citet{draine03}, which is based on an analytical approximation of the  cross-section for 
the dust mixture described in \citet{wd01}.
This approach was used in the analysis of the X-ray rings produced by other GRBs (e.g., \citealt{tm06}), as well as for the \ixpe\ observations of \grb\ \citep{negro23}. 
However, the approximation
may be not accurate enough for our high quality data, especially at the lowest energies (see e.g. fig.\,9 in \citealt{draine03}).

Therefore, we implemented a more appropriate model, dubbed \texttt{ringscat}, based on the exact Mie calculation of the scattering cross-section on spherical grains, adopting several grain compositions and size distributions \citep{zubko04,mathis77}.
This was done as described Appendix~\ref{ringscat},   computing the scattering cross-sections as for the \texttt{xscat} model  \citep{rsmith06} already available within the \texttt{XSPEC} spectral fitting package \citep{arnaud96}.
Based on Eq.~\ref{specmod}, we fit the spectra of each ring
with the \texttt{XSPEC} model: 
\texttt{constant*TBabs*zTBabs*ringscat*pegpwrlw}, where \texttt{ringscat} accounts for the optical depth for different dust models
$\Delta N_{\rm H}\sigma_{\theta_{1,2}}(E)$, the \texttt{constant} is the inverse of the exposure time ($T_{\rm exp}^{-1}$),  and the prompt GRB spectrum, $f(E)$, is modelled as a power law 
(\texttt{pegpwrlw}) modified by absorption \citep{wilms00} both in our Galaxy (\texttt{TBabs}) and in the host galaxy (\texttt{zTBabs}, with redshift fixed at $z=0.151$; \citealt{malesani23}).

\begin{figure}[ht!]
\centering
\includegraphics[width=.37\textwidth, angle=-90,trim=0in 0.8in 0in 0in]{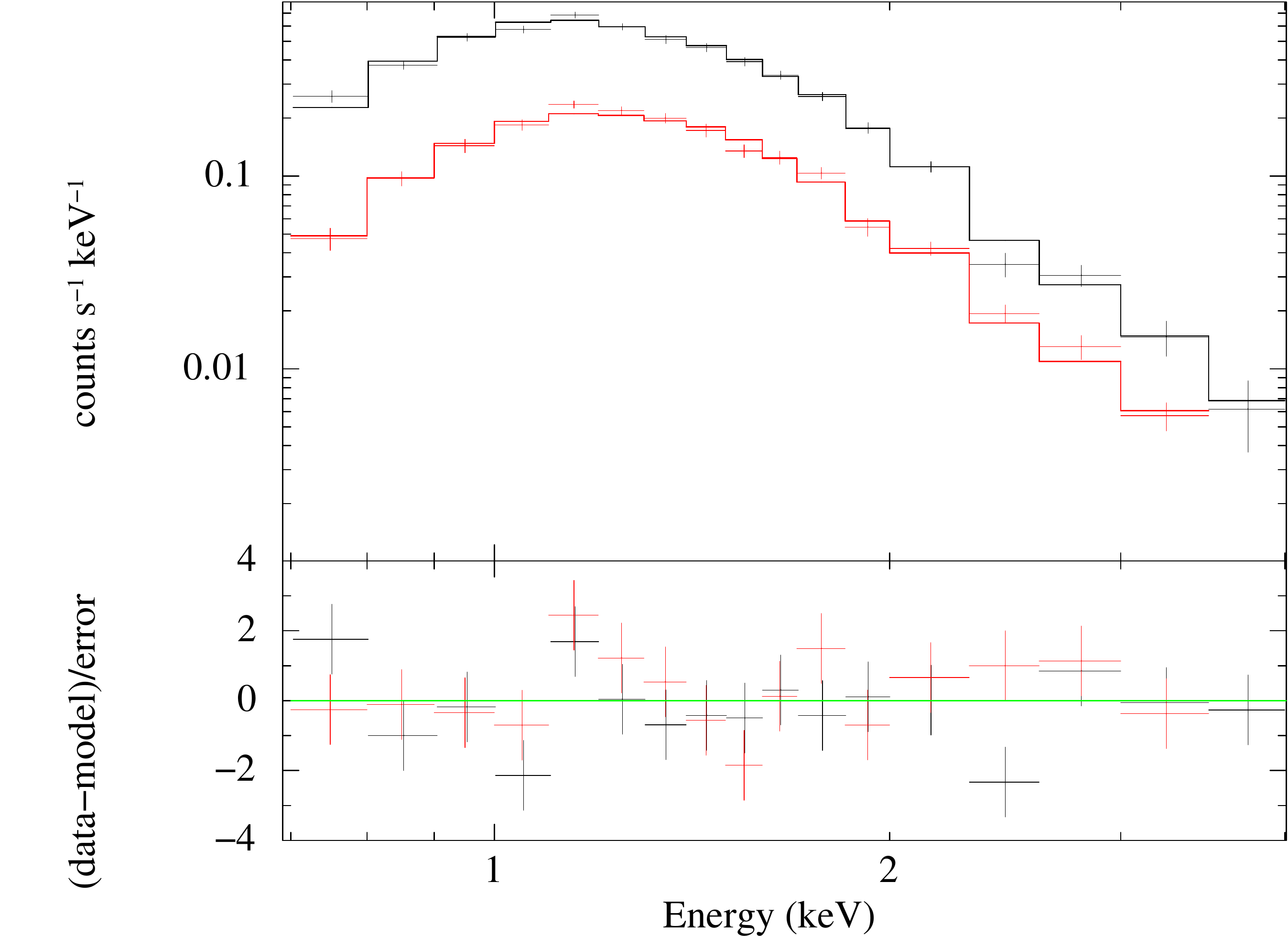}
\includegraphics[width=.37\textwidth, angle=-90,trim=0in 0.8in 0in 0in]{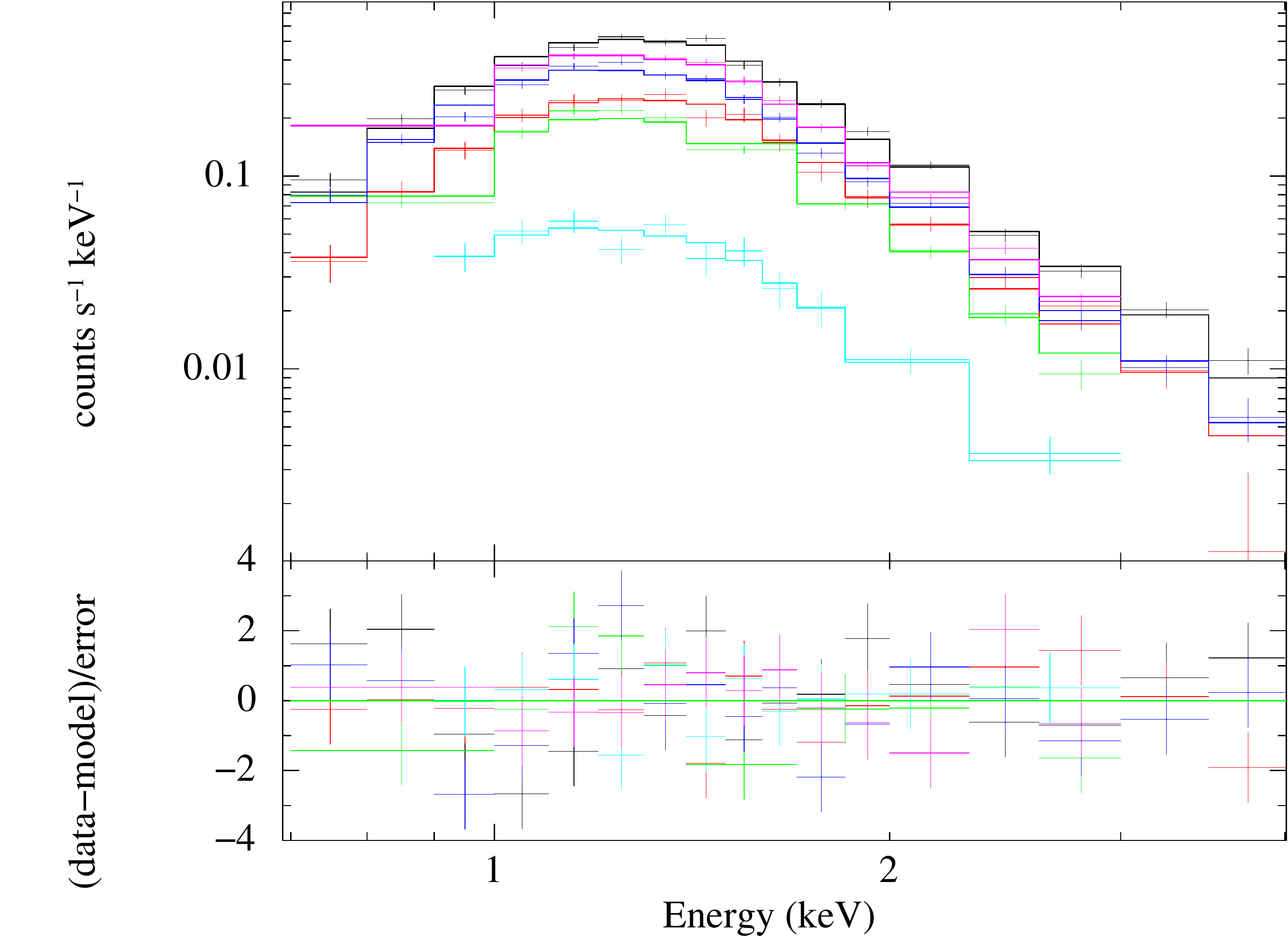}
\includegraphics[width=.37\textwidth, angle=-90,trim=0in 0.8in 0in 0in]{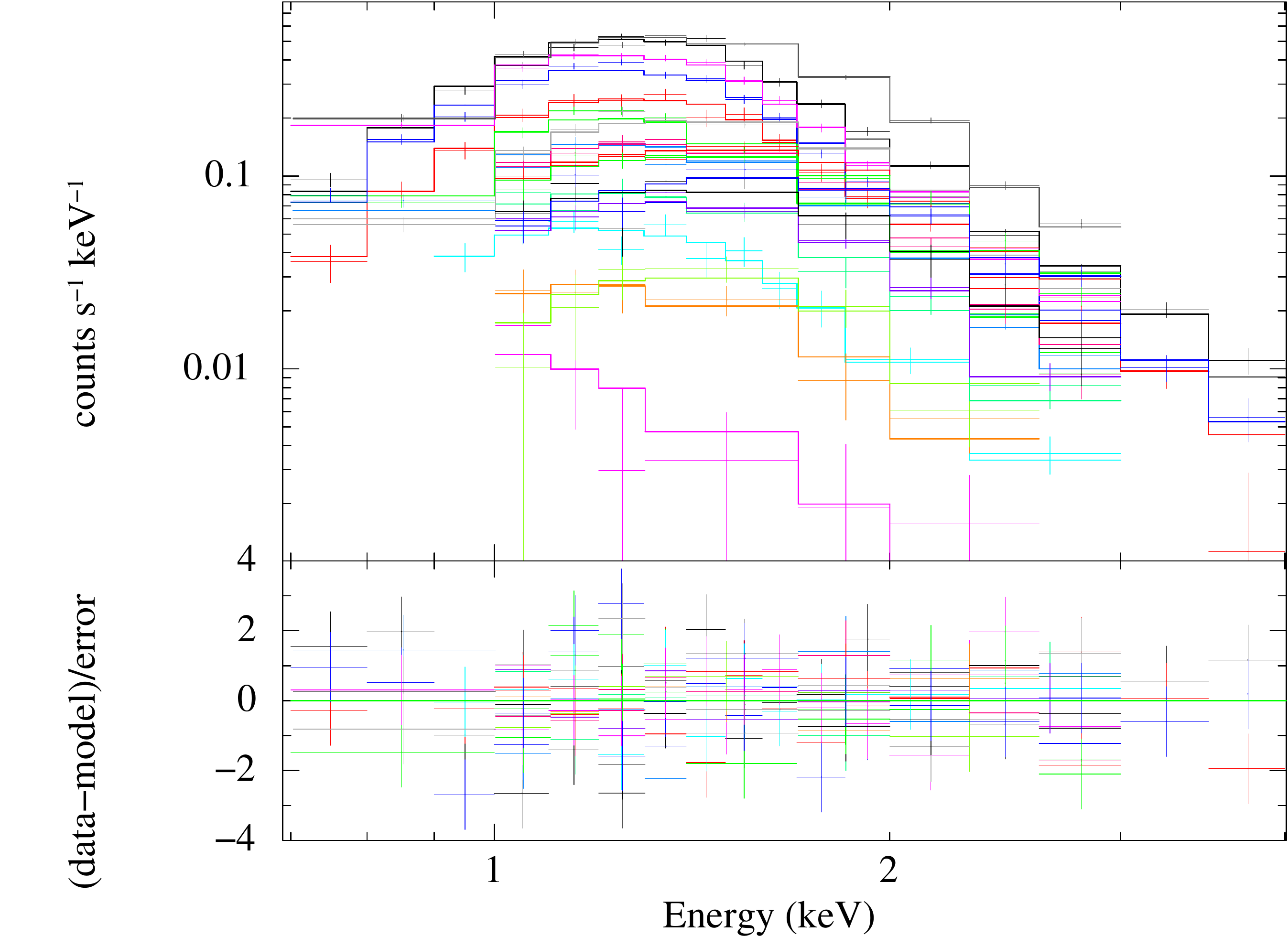}
\caption{Fit of the ring spectra with the BARE-GR-B model (best-fit parameters in Table~\ref{SpecPar}). {\it Top panel:} PN (black) and MOS2 (red) spectra of ring 6 in Obs2. {\it Middle panel:} MOS2 spectra of rings 1--6 in Obs1. {\it Bottom panel:} MOS2 spectra of rings 1--6 in Obs1 and PN spectra of rings 7--19 in Obs2.} 
\label{spectra}
\end{figure}

The ring produced by dust at 0.73 kpc (ring 6) is the brightest one observed in both observations, and the only one for which we  have a reasonable estimate of the quantity of dust producing it.
In fact, from the \citet{lallement22} 3D extinction map, we can assume $\Delta N_{\rm H}=8\times10^{20}$\,cm$^{-2}$ (Appendix~\ref{ism}). 
We  first fit the spectra of ring 6 obtained in Obs2, when it had a radius of $11\farcm5$. This large size reduces the impact of small scale fluctuations in the dust spatial distribution, which cannot be tracked by the low spatial resolution map of \citet{lallement22}.
We  fixed the Galactic absorption at $N_{\rm H,G}=7\times10^{21}$\,cm$^{-2}$, which is the average value of the \citet{planck14} 2D map at the ring position.\footnote{This value was assumed also in all the joint spectral fits of multiple rings, but only for ring 6, while the Galactic absorption for the other rings was let free to vary. We have verified that the resulting best-fit values for each ring are compatible with the $N_{\rm H}$ radial profile derived from \citet{planck14}. 
We note that the statistical uncertainties of most bins in such radial profile is larger than that of the $N_{\rm H,G}$ derived from the spectral fitting of the brightest X-ray rings.}
By simultaneously fitting the PN and MOS2 spectra of Obs2,\footnote{The MOS2 spectrum of Obs1 gave consistent results, except for a $\sim$15\% lower GRB fluence, which can be explained by the fact that dust is not uniformly distributed in this sky area.} we found that all the dust models gave statistically acceptable fits (see an example in the top panel of Fig.\,\ref{spectra}) and yielded similar GRB spectral parameters, with values of the 0.5--5 keV  fluence in the range (2--$4.5)\times10^{-3}$\,erg\,cm$^{-2}$  (see Table~\ref{SpecPar}).

To discriminate between the different dust models and  better constrain the spectral shape of the GRB prompt emission, we  performed a similar analysis by simultaneously fitting the spectra of the other rings.
The rings completely inside the MOS2 field of view in Obs1 (rings 1--6) are produced by dust in the range 350--800\,pc, where a large extinction increase is clearly detected in 3D maps (Appendix~\ref{ism}).
Therefore, based either on \citet{lallement22} or \citealt{green19},  we  can assume $\Delta N_{\rm H}=4\times10^{21}$\,cm$^{-2}$  or $\Delta N_{\rm H}=2.3\times10^{21}$\,cm$^{-2}$, respectively. The fluences derived using the former value are listed in Table~\ref{SpecPar} (obviously, the smaller value of $\Delta N_{\rm H}$  would give $\sim$70\% larger fluences) and the fit with the BARE-GR-B model is shown in the middle panel of Fig.~\ref{spectra}.
 
Finally, we performed a joint fit of the spectra of  rings 1--19 (in addition to the MOS2 spectra, we included also the PN spectra of rings 7--19 during Obs2; see the bottom panel of Fig.~\ref{spectra} for the best-fit model). The results reported in Table~\ref{SpecPar} were obtained assuming $\Delta N_{\rm H}=6.7\times10^{21}$\,cm$^{-2}$ for the cumulative column density in the 19 dust clouds, based on \citet{lallement22}. 
Over this broader distance interval, ranging from 350 pc to the edge of the Galaxy, the discrepancies between the \citet{lallement22} and \citet{green19} maps are smaller, giving only a 15\% larger fluence if the total extinction is derived from the latter map. 

\begin{table*}
\centering                          
\begin{tabular}{lcccccl}
\hline\hline    
Dataset& Dust  & $N_{\rm H,z}$$^a$ & $\Gamma_{\rm GRB}$$^b$ & GRB fluence$^c$ & $\chi^2$/d.o.f. \\   
 & model  & (10$^{21}$ cm$^{-2}$) &  & (10$^{-3}$ erg cm$^{-2}$)\\   
\hline   
Ring 6$^d$ & BARE-GR-B  & 4.9$\pm$0.6  & 1.3$\pm$0.1 & 2.57$\pm$0.05 &36.77/30 \\
& BARE-GR-S &   5.0$\pm$0.6  & 1.2$\pm$0.1 & 2.52$\pm$0.05 & 40.76/30 \\
& BARE-GR-FG &   5.0$\pm$0.6  & 1.2$\pm$0.1 & 2.52$\pm$0.05 & 39.77/30 \\
& COMP-GR-B &   5.8$\pm$0.6  & 1.2$\pm$0.1 & 4.2$\pm$0.1 & 45.55/30 \\
& COMP-GR-S &   5.3$\pm$0.5  & 1.05$\pm$0.09 & 4.4$\pm$0.1 & 34.57/30 \\
& COMP-GR-FG &   5.9$\pm$0.6  & 1.1$\pm$0.1 & 4.0$\pm$0.1 & 41.05/30 \\
& MRN &   6.7$\pm$0.6  & 1.5$\pm$0.1 & 2.03$\pm$0.04 & 44.57/30 \\
& Draine (2003) &   6.7$\pm$0.5  & 1.1$\pm$0.1 & 2.41$\pm$0.06 & 34.94/30 \\
\hline   
Rings 1--6$^e$ &BARE-GR-B &  4.6$\pm$0.4  & 1.39$\pm$0.05 & 1.94$\pm$0.02 & 103.83/72 \\
& BARE-GR-S &   5.3$\pm$0.4  & 1.31$\pm$0.05 & 1.87$\pm$0.02 & 107.46/72 \\
& BARE-GR-FG &   5.6$\pm$0.4  & 1.32$\pm$0.05 & 1.89$\pm$0.02 & 117.45/72 \\
& COMP-GR-B &   5.6$\pm$0.4  & 1.17$\pm$0.05 & 3.09$\pm$0.03 & 110.10/72 \\
& COMP-GR-S &   6.7$\pm$0.4  & 1.08$\pm$0.05 & 3.23$\pm$0.04 & 143.07/72 \\
& COMP-GR-FG &   6.3$\pm$0.5  & 1.09$\pm$0.05 & 2.91$\pm$0.04 & 124.39/72 \\
& MRN &   7.5$\pm$0.5  & 1.44$\pm$0.05 & 1.55$\pm$0.02 & 141.06/72  \\
& Draine (2003) &   7.1$\pm$0.4  & 1.12$\pm$0.05 & 1.59$\pm$0.02 & 99.39/72 \\
\hline
Rings 1--19$^f$ & BARE-GR-B &   4.4$\pm$0.3  & 1.37$\pm$0.04 & 1.67$\pm$0.02 & 194.34/155 \\
& BARE-GR-S &   4.6$\pm$0.4  & 1.20$\pm$0.04 & 1.53$\pm$0.01 & 210.77/155 \\
& BARE-GR-FG &   4.6$\pm$0.3  & 1.18$\pm$0.04 & 1.54$\pm$0.01 & 235.18/155 \\
& COMP-GR-B &   5.1$\pm$0.2  & 1.08$\pm$0.04 & 2.47$\pm$0.03 & 204.99/155 \\
& COMP-GR-S &   5.2$\pm$0.2  & 0.86$\pm$0.04 & 2.52$\pm$0.03 & 298.95/155 \\
& COMP-GR-FG &   5.0$\pm$0.4  & 0.89$\pm$0.04 & 2.28$\pm$0.02 & 257.26/155 \\
& MRN &   6.0$\pm$0.4  & 1.21$\pm$0.04 & 1.24$\pm$0.01 & 302.54/155  \\
& Draine (2003) &   6.3$\pm$0.4  & 1.00$\pm$0.04 & 1.25$\pm$0.01 & 211.10/155 \\
\hline                                  
\end{tabular}\\
\footnotesize{$^a$Absorption in the host galaxy ($z=0.151$). $^b$Photon index of the GRB prompt emission $^c$Unabsorbed GRB fluence in the 0.5--5 keV energy band. $^d$PN and MOS2 spectra of ring 6 in Obs2, assuming $\Delta N_{\rm H}$=8$\times$10$^{20}$ cm$^{-2}$ in the dust cloud. $^e$MOS2 spectra of rings 1--6 in Obs1, assuming $\Delta N_{\rm H}=4\times10^{21}$\,cm$^{-2}$ in the 6 dust clouds. $^f$MOS2 spectra of rings 1--6 in Obs1 and PN spectra of rings 7--19 in Obs2, assuming $\Delta N_{\rm H}=6.7\times10^{21}$\,cm$^{-2}$ in the 19 dust clouds. }\\
\caption{Best-fit parameters of the GRB prompt emission derived from different choices of dust-scattering rings and dust models. The Galactic absorption of ring 6 is fixed at $N_{\rm H,G}$=7$\times$10$^{21}$ cm$^{-2}$. All uncertainties are at 1 $\sigma$. \label{SpecPar}}
\end{table*}

\subsection{Properties of the GRB prompt emission}

The dust model that provides the best fit (null-hypothesis probability of 0.02 for the joint fit of 19 rings; Fig.\,\ref{spectra}) is BARE-GR-B \citep{zubko04}, which also resulted as the best-fitting model in previous studies of dust-scattering X-ray halos (e.g., \citealt{rsmith06,tiengo10}). 
However, also COMP-GR-B, BARE-GR-S and the \citet{draine03} models give a reasonably good fit to the full set of spectra (Table~\ref{SpecPar}). According to these models, the power-law photon index of the GRB prompt emission ranges from $\sim$1.0 to $\sim$1.4, with the steepest slope (1.37$\pm$0.04) obtained with BARE-GR-B. 

If we exclude the \citet{draine03} model, which is not accurate enough at low energies,
the hydrogen column density in the host galaxy can be constrained within the narrow range $N_{\rm H,z}=(4.1$--$5.3)\times10^{21}$\,cm$^{-2}$   (Table~\ref{SpecPar}).   However, this result depends on the value of the absorption in our own Galaxy, which we fixed to $N_{\rm H,G}=7\times10^{21}$\,cm$^{-2}$ for ring 6. To evaluate the impact of a different assumption on the Galactic absorption, we fixed it at $N_{\rm H,G}=5.38\times10^{21}$\,cm$^{-2}$ \citep{willingale13}, obtaining for the BARE-GR-B model a slightly worse fit ($\chi^2$ from 194.34 to 200.08 for 155 d.o.f.) and an increase of   $N_{\rm H,z}$ from $(4.4\pm0.3)\times10^{21}$\,cm$^{-2}$ to $(6.6\pm0.4)\times10^{21}$\,cm$^{-2}$. 
Assuming instead $N_{\rm H,G}=9\times10^{21}$\,cm$^{-2}$, which is the largest value displayed by the \citet{planck14} map within the region covered by the X-ray rings (Fig.\,\ref{fig:planck}), we obtain a better fit ($\chi^2$=188.55 for 155 d.o.f.) and an intrinsic absorption of $N_{\rm H,z}=(1.8\pm0.2)\times10^{21}$\,cm$^{-2}$.

Our estimate of the GRB fluence depends on our  assumptions on the column density of dust in the clouds, based on 3D extinction maps (Appendix~\ref{ism}). First of all, we note that the fluence values for each dust model are systematically lower when a larger number of rings is considered (Table~\ref{SpecPar}). This effect can be explained by the fact that the extinction excess derived from the maps
includes also the contribution from diffuse (i.e. dust not concentrated in the layers associated to the rings) dust or unresolved scattering rings and should therefore be considered as an upper limit to the amount of dust in the thin layers generating the X-ray rings.
The most conservative lower limit to the GRB fluence can therefore be derived from the
fit of the spectra of the 19 rings
with the \citet{draine03} model,\footnote{The same fluence is obtained also with the MRN model \citep{mathis77}, which however provides a very poor fit to the 19 spectra (null-hypothesis probability of 10$^{-11}$).}
which gives a GRB fluence of $1.25\times10^{-3}$\,erg\,cm$^{-2}$.

Excluding the COMP-GR-S model,
which does not adequately fit the ring spectra, 
the largest fluence is derived adopting 
the COMP-GR-B model. Further increasing the value reported in Table~\ref{SpecPar} by assuming the extinction in the 6 dust clouds from \citep{green19} and by considering
that $\sim$30\% of the dust between 350 and 800 pc might not be inside the 6 clouds, we set $7\times10^{-3}$\,erg\,cm$^{-2}$ as the most conservative upper limit to the 0.5--5 keV fluence of \grb.

The most reliable estimate for the GRB fluence is that obtained by fitting 
the spectrum of ring 6 with the BARE-GR-B model ($2.6\times10^{-3}$\,erg\,cm$^{-2}$).
However, also this values is affected by 
unavoidable systematic uncertainties, which, although difficult to quantify, are likely within the lower and upper limits
derived above. First of all, it is based on an excess of optical extinction detected (at the correct distance of 0.73 kpc) in a map with poor angular resolution \citep{lallement22}, while X-ray data exhibit clear evidence for small scale variations of the dust distribution in this cloud (Fig.\,\ref{fig:azimuth}). Moreover, \xmm\ can resolve two rings (ring 5 and 6) within the peak in the \citet{lallement22} map, but their relative intensity (ring 5 is about 3 times fainter than ring 6) is significantly different in the two observations. Since this indicates that these two dust clouds might have substantially different shape and size, we attributed the whole extinction excess to the dust producing the brightest ring (ring 6). 
Finally, our fluence estimates depend on the conversion from $\Delta$A$_V$ to $\Delta N_{\rm H}$, which we have derived for our sky area from \citet{planck14} and \citet{schlafly11} (Appendix~\ref{ism}). This relation is also affected by non-negligible systematic uncertainties, as demonstrated by the scatter displayed by the measures reported in previous studies, covering different sky directions and astrophysical objects (e.g., \citealt{zhu17}).\\

\section{Discussion}

The large collecting area and wide field of view of \xmm\ allowed us to detect  20  dust scattering rings around the extremely bright \grb. They were produced by dust concentrations at distances between 300 pc and 18.6 kpc.  Besides deriving accurate measurements of the distance of the dust layers from the time delay and position of the scattered photons
(relative precision ranging from 0.05\% to 0.8\%; Table~\ref{DistTab}), we could 
constrain the prompt X-ray emission of the burst through a detailed spectral analysis.

Analysis of the \grb\ dust scattering rings  were also reported by \citet{vasilopoulos23} and \citet{williams23}, who used \swi/XRT,  and by  \citet{negro23}, who used \ixpe\ data. Besides being based on data with a higher counting statistics, our analysis does not rely on a single model for the dust grain size distribution and composition. The use of different dust models \citep{draine03,zubko04,mathis77} allowed us to exclude some of them, including the MRN model (Table~\ref{SpecPar}), which adequately fit the \sw/XRT data \citep{williams23},  and to evaluate their effect
on the reconstruction of the GRB soft X-ray prompt emission. 

The dust models were implemented using the exact Mie computations for the scattering process for spherical dust grains (Appendix~\ref{ringscat}), which 
is valid in the whole energy range where the X-ray rings are detected (0.7--4 keV). In fact, the vast majority of the rings signal in \xmm\ has been detected below 2 keV, where 
X-ray scattering models based on the Rayleigh-Gans approximation are inadequate for the analysis of our high quality spectra.

Our best estimate of the X-ray fluence of \grb\ in the 0.5--5 keV band
is $2.6\times10^{-3}$\,erg\,cm$^{-2}$, which, considering systematic uncertainties, is consistent with the values obtained by extrapolating to lower energies the model of the prompt emission observed by \fermi/GBM ($1.7\times10^{-3}$\,erg\,cm$^{-2}$; Lesage et al. in prep.).\footnote{A 0.5--5 fluence of 5$^{+10}_{-2}\times10^{-5}$\,erg\,cm$^{-2}$ is instead obtained by extrapolating the 80--800 keV fluence observed by GRBAlpha assuming the spectrum observed during the first 4 s of the main GRB peak \citep{ripa23}. This fluence is much lower than that estimated from \fermi/GBM data, which, however, is based on the spectrum observed during the bulk of the GRB prompt emission, over a broader energy band (8 keV -- 1 MeV), and corrected for saturation effects.}
As a matter of fact, considering all the dust models producing acceptable fits to the ring spectra (Table~\ref{SpecPar}) and different proxies for the quantity of dust producing them (Appendix~\ref{ism}), our analysis 
safely constrains the 0.5--5 keV fluence of \grb\ between $10^{-3}$ and $7\times10^{-3}$\,erg\,cm$^{-2}$ 
(corresponding to $2.4\times10^{-3}$ to $1.3\times10^{-2}$\,erg\,cm$^{-2}$ in the 1--10 keV range). 

A lower 1--10 keV fluence of (1.6--$6.1)\times10^{-4}$\,erg\,cm$^{-2}$ was instead estimated from the analysis of the dust rings observed with \ixpe, which allowed \citet{negro23} to derive limits on the X-ray polarization  of \grb. 
During the \ixpe\ observations, the  rings produced by dust closer than 2 kpc were outside the field of view and, due to the limited angular resolution and collecting area, only two rings corresponding to broad dust distributions centered at 3.75 and 14.41 kpc could be identified.  
The GRB fluence was then reconstructed using the \citet{draine03} model, assuming that all the optical extinction at distances greater than 3 kpc is due to two narrow dust layers at such distances \citep{negro23}. As demonstrated by the complex dust distribution unveiled by our analysis of the \xmm\ data, this is an overestimate of the quantity of dust in these dust layer, leading to a lower estimate of the GRB fluence. Moreover, as already noted, the fluences derived from the \citet{draine03} model are smaller than those obtained from the other dust models and, although appropriate above 2 keV, where \ixpe\ operates, this model cannot be safely used to extrapolate the ring spectra at lower energies.

It is important to stress that, contrary to the case of the fluences, the spectral slopes we derived for the GRB prompt emission do not depend on the assumed quantity of dust producing the rings. They depend only on the shape of the scattering cross section, in turn related to the composition, shape, and size distribution of the grains. Considering only the dust models giving acceptable fits to the 19 complete rings detected by \xmm, the power-law photon index spans from 1 to 1.4. The steepest spectrum ($\Gamma_{\rm GRB}=1.37\pm0.04$) is derived from the best-fitting model (BARE-GR-B), whereas a harder slope is suggested by the analysis of the Konus-Wind ($\Gamma_{\rm GRB}=1.09\pm0.01$ in the first 20 s of the main peak; \citealt{frederiks22}), GRBAlpha ($\Gamma_{\rm GRB}=0.7\pm0.1$ in the first 4 s of the main peak; \citealt{ripa23}) and \fermi/GBM ($\Gamma_{\rm GRB}\sim0.4$; Lesage et al. in prep.) spectrum at the lowest energies. This might indicate a soft excess in the prompt emission of \grb. The broadening of the typical synchrotron spectrum responsible for the main spectral peak is predicted in the numerical simulations of the lepto-hadronic models for GRBs. For example, \citet{rudolph22} showed that a broad flat spectrum due to the synchrotron emission of secondary pairs from $\gamma\gamma$-annihilation may significantly affect the spectral slopes below the spectral peak. 
The indication of a soft excess derived from our analysis is 
at odds with what derived in other GRBs for which prompt soft X-ray data are available \citep{oganesyan17,oganesyan18}. Indeed, our measured fluence and spectral slope  exclude the existence of a marked flattening of the prompt emission spectrum of \grb\ below a few keV or tens of keV, unless it is balanced by an additional component in the soft energy range.

The possibility to fit the scattered emission with a physically motivated model also in the soft X-ray band, made it possible to constrain the  absorption 
in the host galaxy in the range $N_{\rm H,z}=(4.1$--$5.3)\times10^{21}$\,cm$^{-2}$ (to be extended by $\sim$50\%, if the systematic uncertainty on Galactic absorption is taken into account). These values are consistent with the intrinsic absorption $N_{\rm H,z}=(1.4\pm0.4)\times10^{22}$\,cm$^{-2}$ derived from the spectral fit of the afterglow observed with \swi/XRT \citep{williams23}, especially if we consider that it was obtained by assuming $N_{\rm H,G}$=5.38$\times10^{21}$\,cm$^{-2}$ \citep{willingale13}, whereas a larger Galactic absorption ($N_{\rm H,G}\sim9\times10^{21}$\,cm$^{-2}$) at the afterglow position is indicated by the \citet{planck14} map (Fig.~\ref{fig:planck}).
The brightness of the \grb\ X-ray afterglow makes it possible to study the evolution of the local absorption with time (Campana et al. in prep.) and therefore our constraints on the absorption of the prompt emission are fundamental to understand how the interstellar medium of the host galaxy was affected by the propagation of the GRB radiation.

Further studies of these extraordinary X-ray rings, including a detailed analysis of both the radial and azimuthal spectral variability of individual rings, will allow us to characterise better the Galactic interstellar medium, reducing the systematic uncertainties affecting the reconstruction of the \grb\ soft X-ray spectrum and the absorption in the host Galaxy. Furthermore, a detailed modelling, including also the different quantity of dust encountered by X-rays as the rings expand, is necessary to estimate the contamination of the X-ray afterglow by dust-scattered X-ray---especially at early times, when the ring size was smaller and \grb\ was observed with operating modes without full imaging capabilities.\\





\begin{acknowledgments}
The scientific results reported in this article are based on 
observations obtained with \xmm, an ESA science mission with 
instruments and contributions directly funded by ESA Member States and NASA.
AT, AS, PE, SM and MR acknowledge financial support from the Italian Ministry for University and Research, through grant 2017LJ39LM (UNIAM). ŽB and VJ acknowledge support by the Croatian Science Foundation for a project IP-2018-01-2889 (LowFreqCRO). AB acknowledges support from the European Research Council through the Advanced Grant MIST (FP7/2017-2022, No.742719).
\end{acknowledgments}

%

\facilities{\xmm}


\software{SAS
(v19.1.0; \citealt{gabriel04}), HEASoft package (v.6.31;
\citealt{ftools14}), Xspec
(v12.13.0c; \citealt{arnaud96}), xscat (v1.0.0; \citealt{smith16}), dustmaps (v1.0.4; \citealt{green18}). }




\appendix
\section{X-ray ring expansion and azimuthal variability} \label{images}

\begin{figure*}[h!]
    \centering
    \includegraphics[width=.32\textwidth]{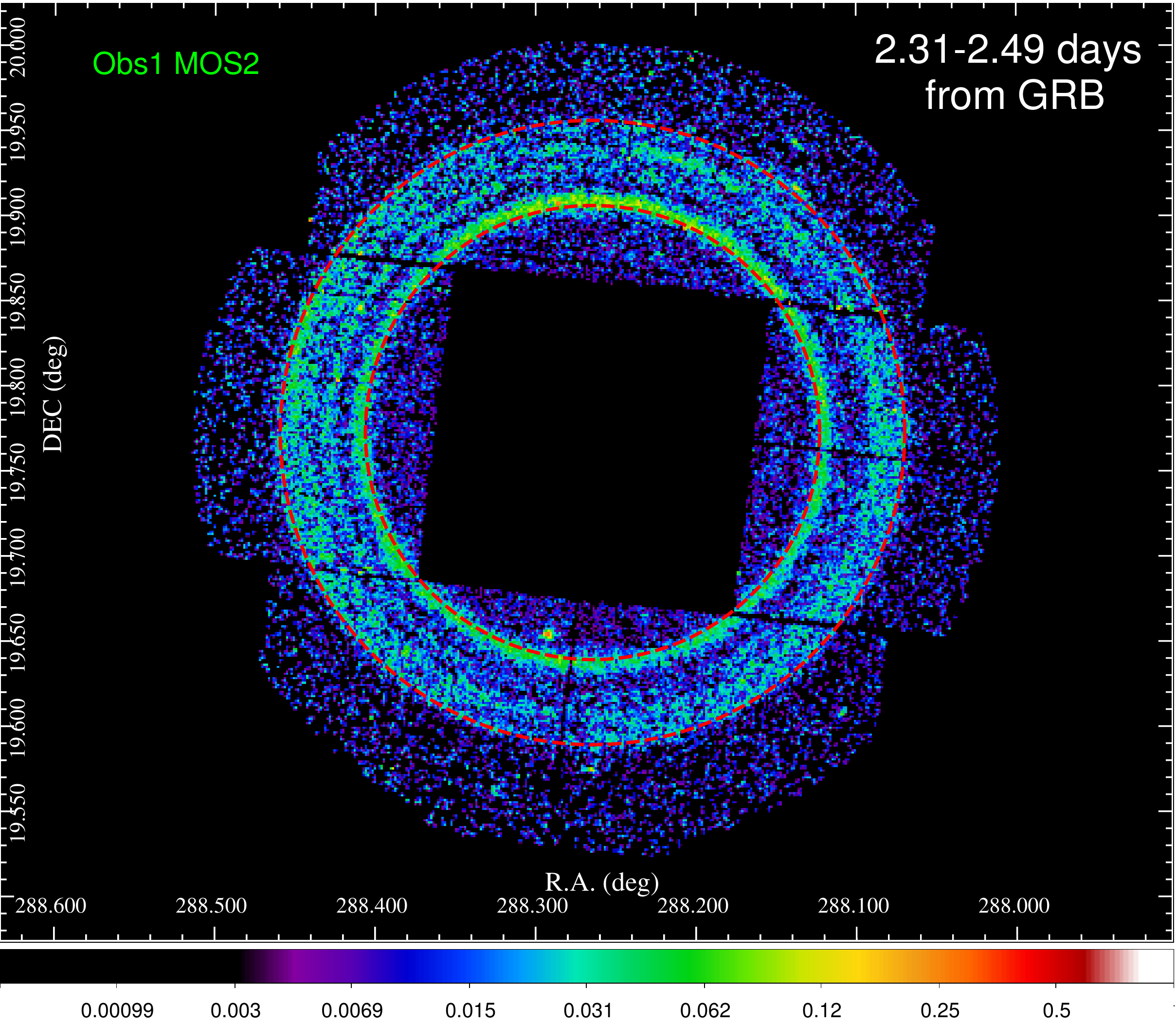}
    \includegraphics[width=.32\textwidth]{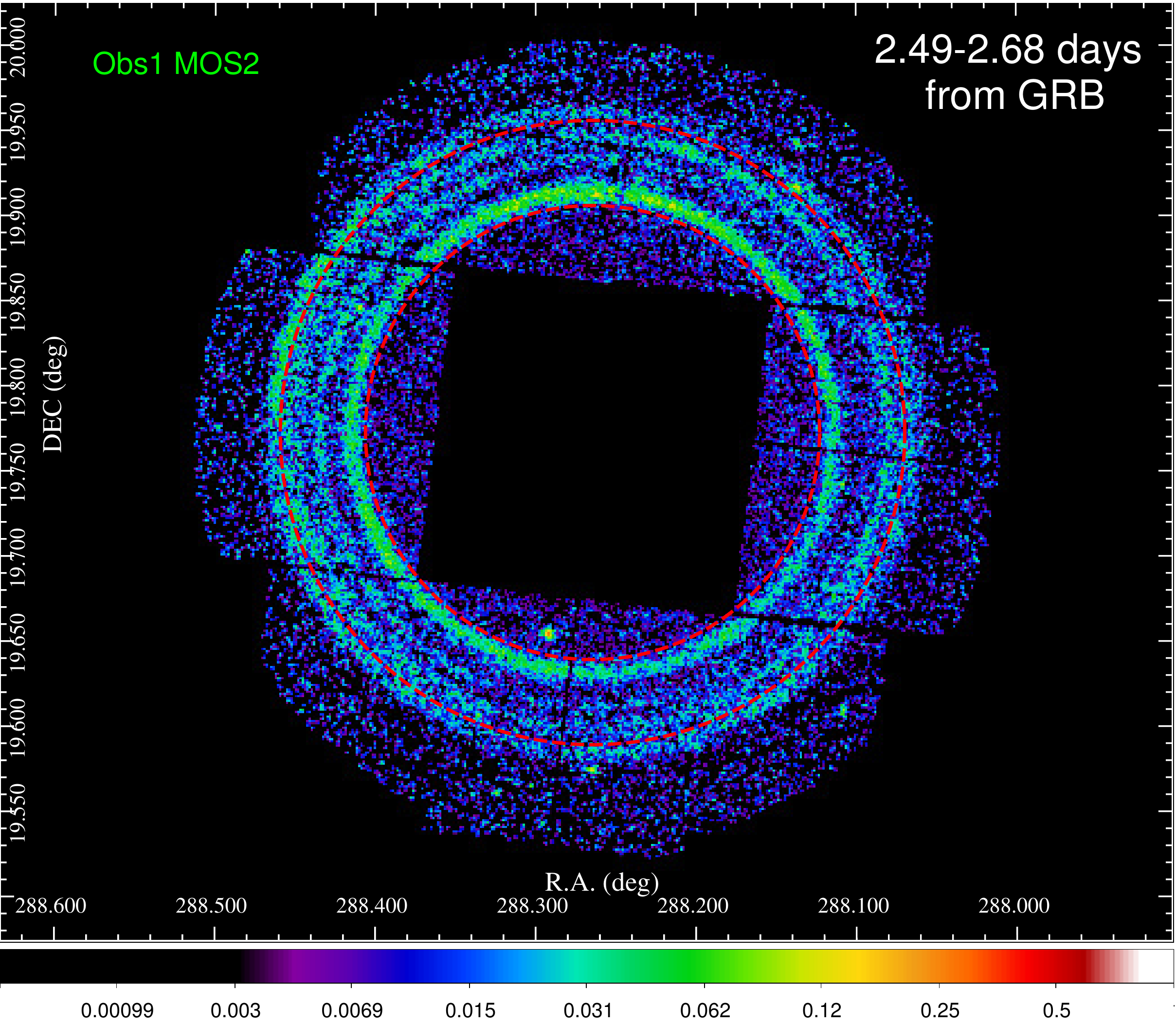}
    \includegraphics[width=.32\textwidth]{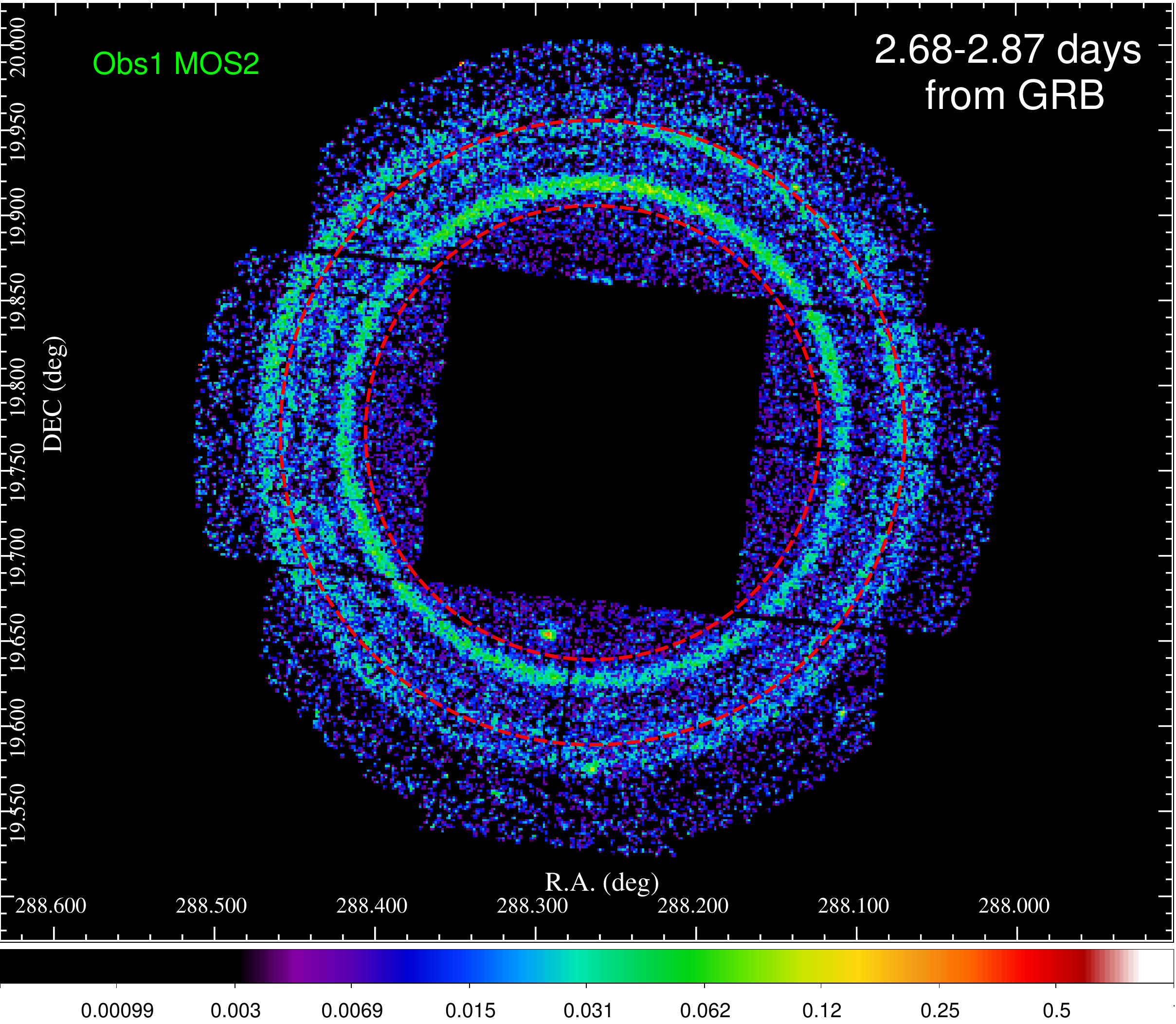}
    \includegraphics[width=.32\textwidth]{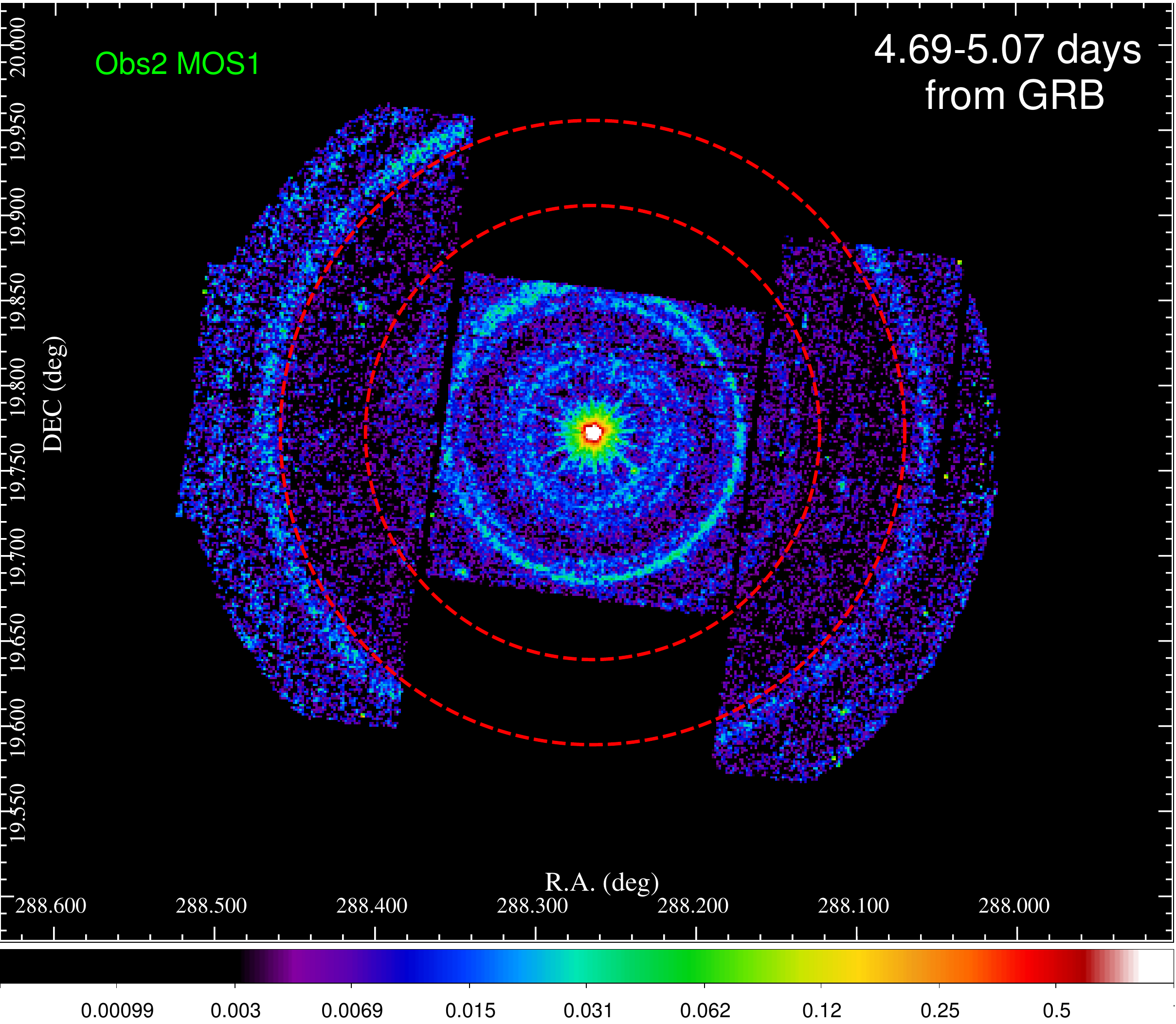}
    \includegraphics[width=.32\textwidth]{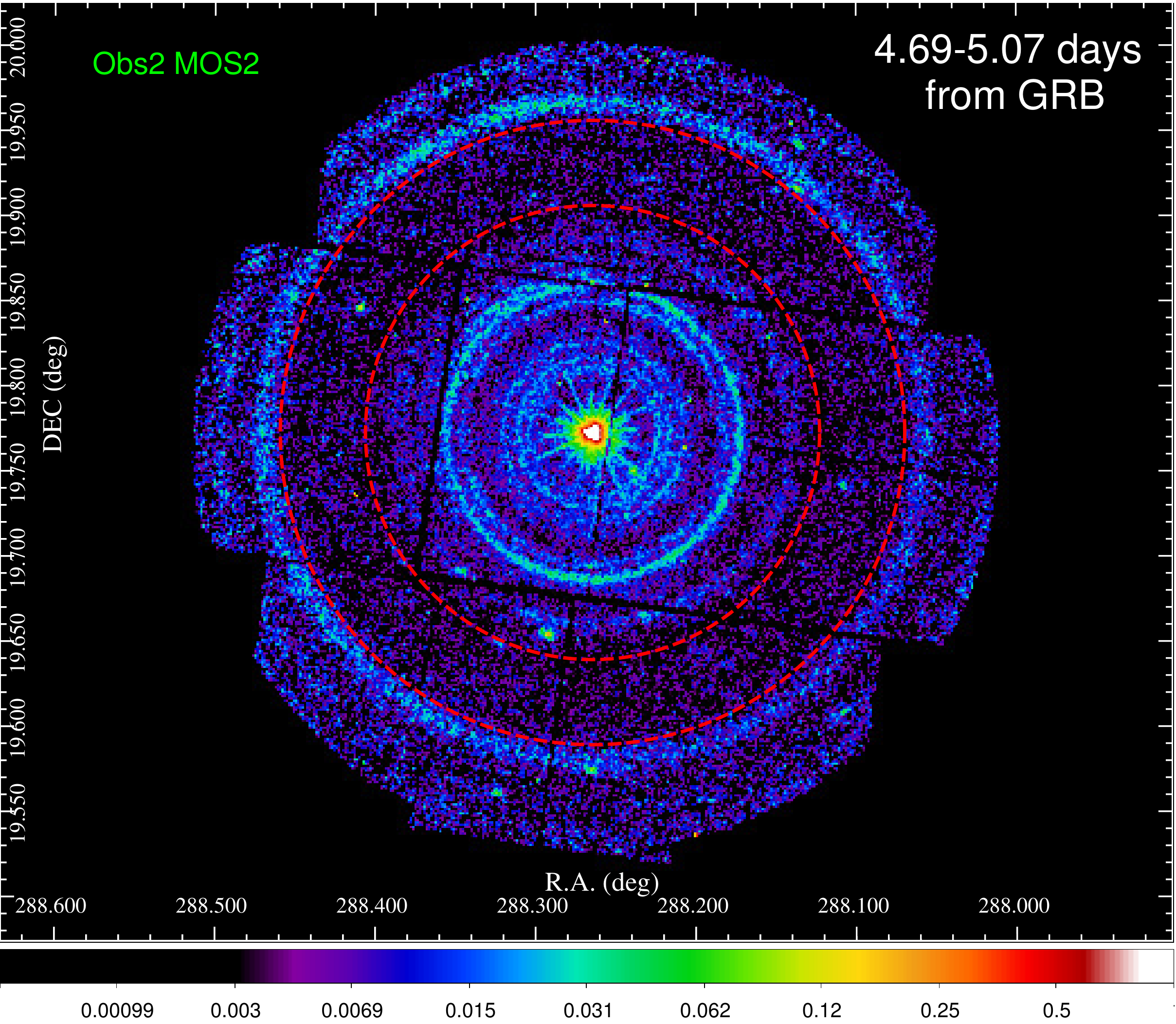}
    \includegraphics[width=.32\textwidth]{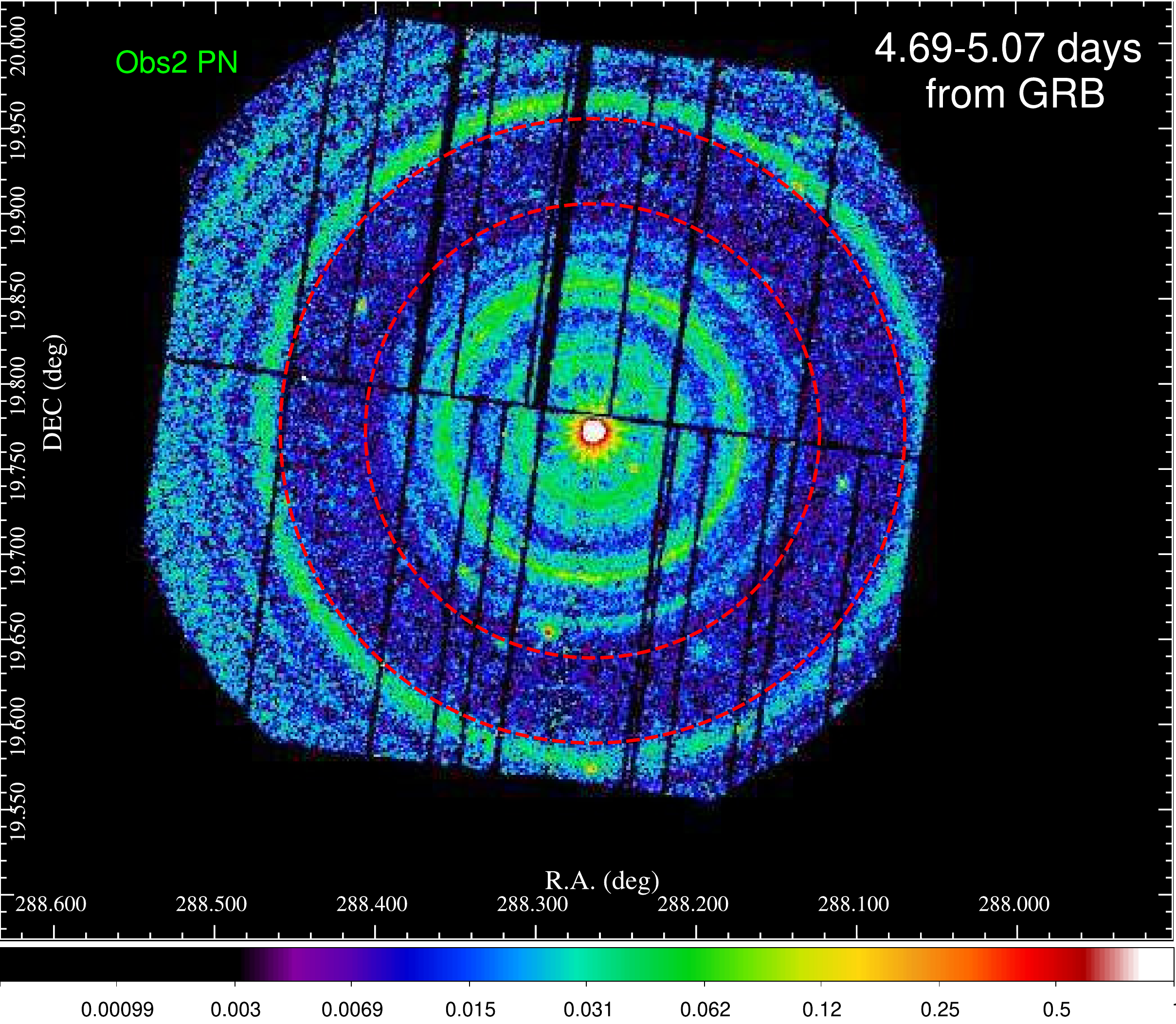}
\caption{EPIC exposure-corrected 0.7--4 keV images, in units of counts s$^{-1}$ arcmin$^{-2}$, of the expanding rings from Obs1 (top panels: MOS2 data in three consecutive time intervals with $\sim$16 ks of exposure time each) and Obs2 (bottom panels, from left to right: MOS1, MOS2 and PN data for the full 33 ks time interval of quiescent background). All the images have been smoothed with a Gaussian kernel of $\sigma=2\farcs5$.  Two red circles of radii 8$^{\prime}$ and 11$^{\prime}$ are shown as a reference for ring expansion.}
    \label{fig:ringsxmm}
\end{figure*}

The top panels of Fig.\,\ref{fig:ringsxmm} show the MOS2 images obtained by dividing the first \xmm\ observation in three time intervals, each of $\sim$16 ks exposure. More individual X-ray rings can be identified with respect to Fig.\,\ref{fig:ringsmos2}, where some of them were superimposed due to their expansion during the observation. Such expansion can be evaluated from the comparison with the reference red circles (8$^{\prime}$ and 11$^{\prime}$ radii). The same red circles are displayed also in the bottom panels, where the corresponding images for the second \xmm\ observation are shown for MOS1, MOS2 and PN. 
In this observation, performed more than 3 days later, the bright ring that was close to the inner red circle in Obs1 (actually comprising ring 5 and 6),
had moved beyond the outer reference circle. We have verified that the expansion rate of ring 6, whose radius can be precisely measured in time-resolved images of both \xmm\ observations, is consistent with an origin within $\sim$1000 seconds from the main GRB event.

The rings detected at larger radii in Obs1 (rings 0--4) were already at least partly outside the field of view of all the EPIC cameras in Obs2. The corresponding MOS1 image shows how this instrument could observe complete rings only with its central CCD, which was  not operated in 2D imaging mode in Obs1. 

\begin{figure}[h!]
    \centering
    \includegraphics[width=0.8\hsize]{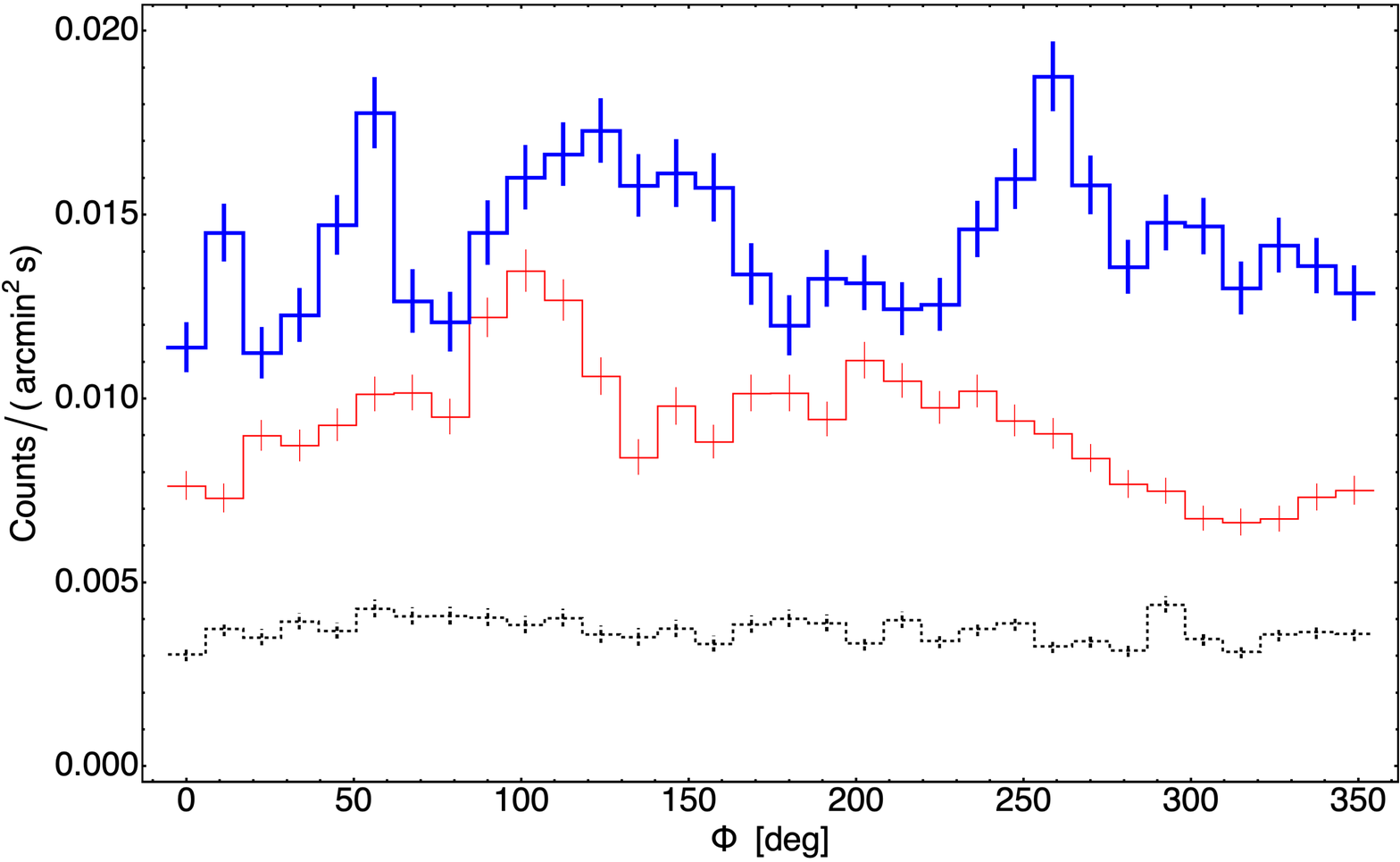}
\caption{Azimuthal distribution of the MOS2 exposure-corrected surface brightness in the 0.7--4 keV band detected between $4\farcm8$ and $5\farcm8$ (blue thick solid line) and between $11\farcm1$ and $12\farcm1$ (red thin solid line) from the GRB position in Obs2. Each annulus contains two rings: 13 and 14 produced by dust at $\sim$3.5 kpc, and rings 5 and 6, produced by dust at $\sim$0.7 kpc.
Both annuli are compared to the same background signal (black dashed line), extracted from a 8$^{\prime}$--$10\farcm5$ annulus, where only two very faint rings (7 and 8) have been detected. All point sources have been excluded.} 
    \label{fig:azimuth}
\end{figure}

Partially imaged rings were not considered for the spectral analysis, in order to reduce the bias in the reconstruction of the GRB fluence caused by the marked azimuthal variation observed in the ring intensity.  
As an example, Fig.\,\ref{fig:azimuth} shows the azimuthal distribution of the surface brightness of the brightest X-ray rings in Obs2. The presence of adjacent bins with statistically incompatible surface brightness indicates significant variations on spatial scales as small as 1$^{\prime}$. 

\section{The dust-scattering spectral model} \label{ringscat}

To reconstruct the GRB prompt emission from the spectrum of each ring (Eq.~\ref{specmod}), we need a model for the optical depth, $\Delta N _{\rm H}\sigma_{\theta_{1,2}}(E)$, for single scattering between the angles corresponding to the inner ($\theta_1$) and outer ($\theta_2$) ring radius, for a populations of dust grains with column density $\Delta N _{\rm H}$. We have therefore implemented a new \texttt{XSPEC} multiplicative model, called \texttt{ringscat}, whose input parameters are $\Delta N _{\rm H}$, $\theta_1$, $\theta_2$, and an integer number to identify different models for dust composition and grain size distribution.

To compute the scattering cross-section, we took advantage of the publicly available software\footnote{\url{https://github.com/AtomDB/xscat}} developed to produce the \texttt{XSPEC} \texttt{xscat} model \citep{smith16}. This extinction model is based on the cross-section for scattering at angles greater than $\theta$, $\sigma_\mathrm{\theta}(E)$,  computed using the  exact Mie theory applied to a population of spherical grains. The cross-section in \texttt{ringscat} can then be simply calculated as:

\begin{equation}
\sigma_{\theta_{1,2}}(E)=\sigma_{\theta_1}(E)-\sigma_{\theta_2}(E). 
\label{sigma}
\end{equation}
In particular, we computed $\sigma_\mathrm{\theta}(E)$ 
in the 0.4--4 keV energy range (with a resolution of 30 eV), for 38 angles
between 2$^{\prime}$ and 12$^{\prime}$, 
for the following dust models: BARE-GR-B, BARE-GR-S, BARE-GR-FG, COMP-GR-B, COMP-GR-S, COMP-GR-FG \citep{zubko04}, and MRN \citep{mathis77}.

\section{Multi-wavelength constraints on the Galactic interstellar medium towards \grb} \label{ism}

To derive the GRB fluence from the X-ray spectrum of a dust-scattering ring, we need an independent estimate of the quantity of dust in the corresponding dust cloud. Similarly, to constrain the amount of absorption in the host galaxy, we need to assume the value of the Galactic absorption in the direction of the X-ray rings. The latter information can be derived from 2D reddening maps (e.g., \citealt{schlafly11,planck14}), whereas 3D maps (e.g., \citealt{green19,lallement22}) are required to evaluate the individual contribution of each dust cloud. 
The \citet{lallement22} map covers a 6$\times$6$\times$0.8 kpc$^3$ volume with a resolution of 25 pc. The \citet{green19} data are instead  defined on 120 distance bins logarithmically spaced  in distances from 63 pc to 63 kpc, with angular sightlines of a typical scale ranging from $3\farcm4$ to $13\farcm7$.

\begin{figure}
    \includegraphics[width=0.5\textwidth]{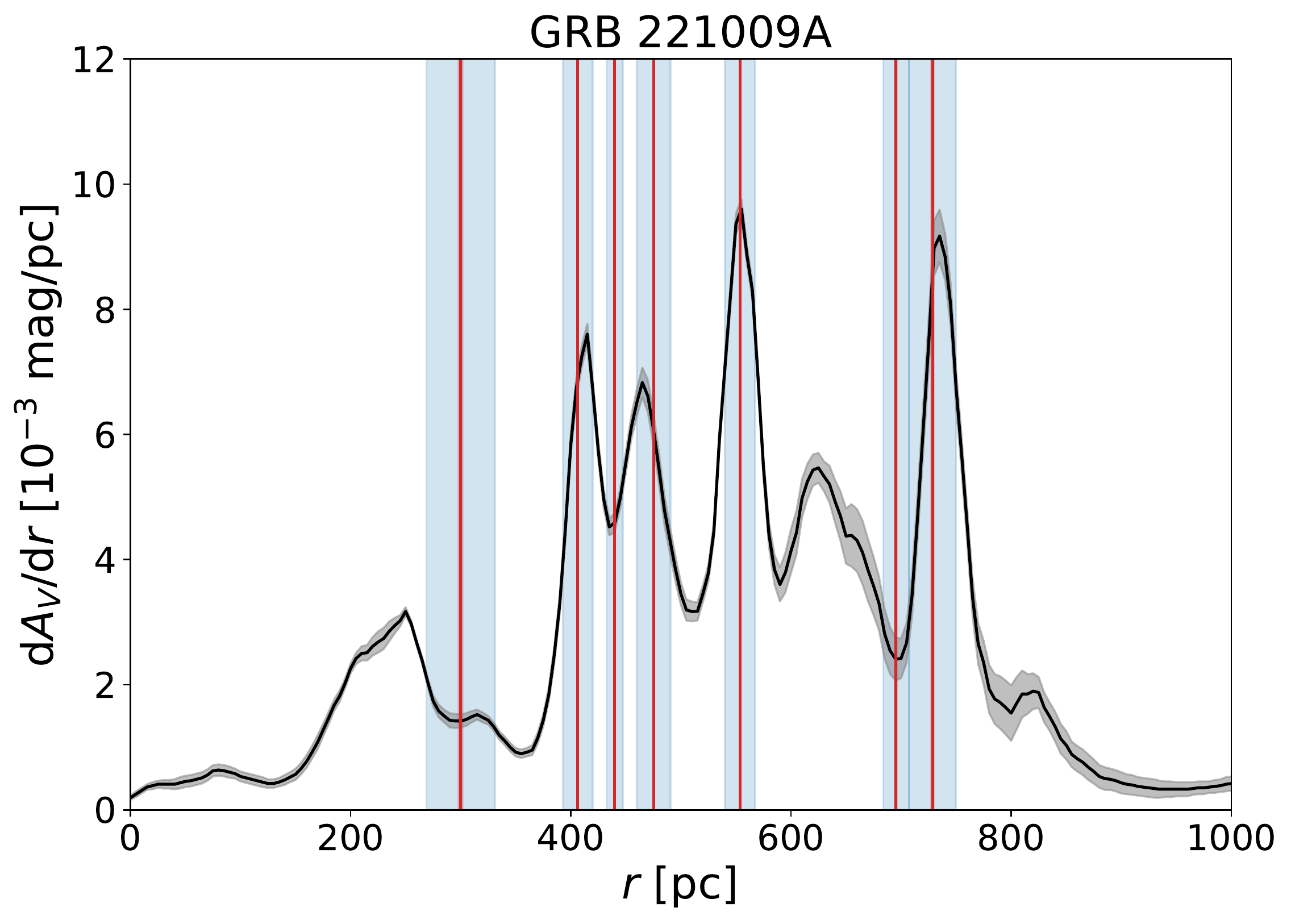}
    \includegraphics[width=0.5\textwidth]{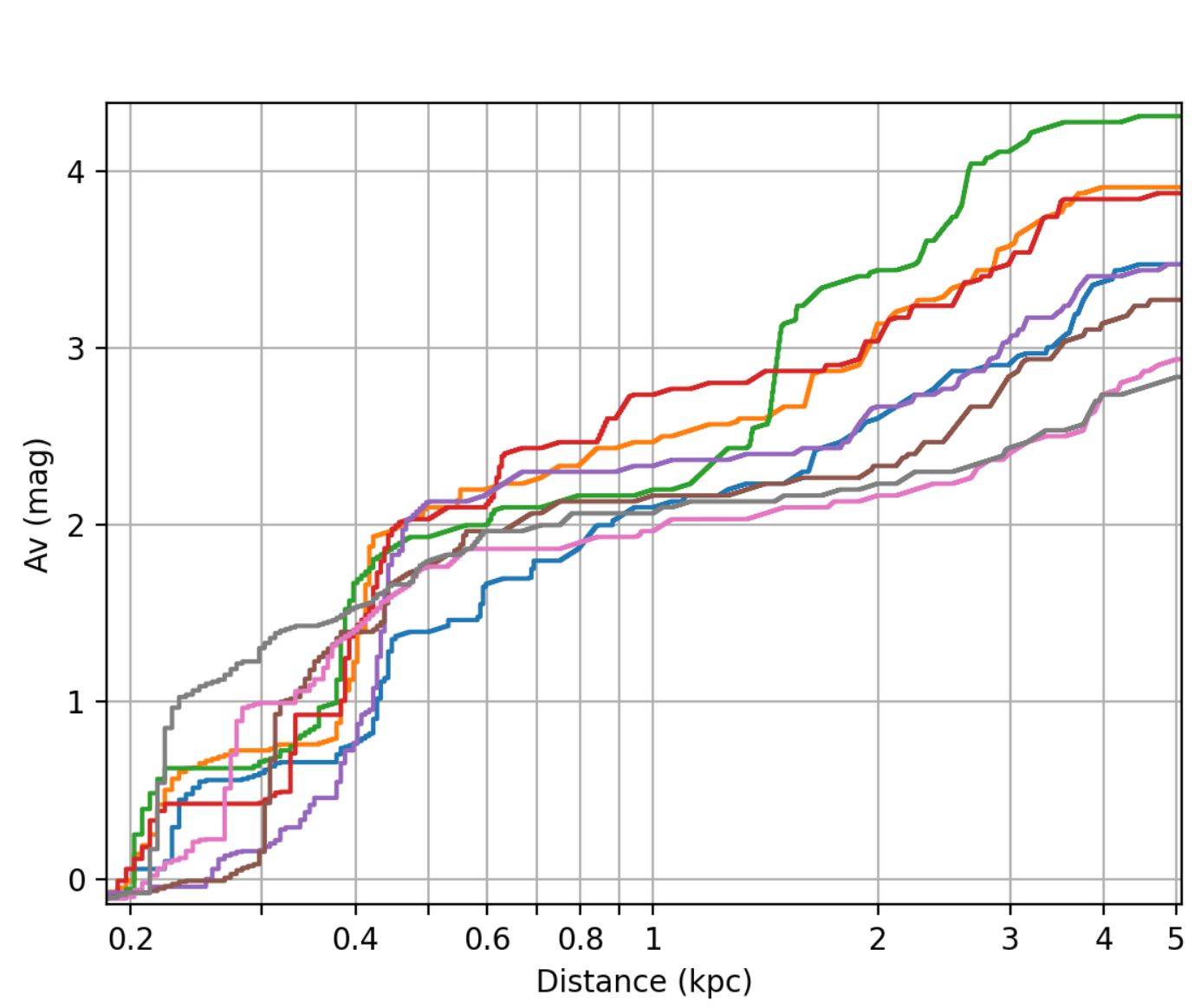}
\caption{ {\it Left panel:} Differential extinction in the direction of \grb\ from \citet{lallement22}. Vertical red lines and light blue shaded areas indicate the distances and widths of dust layers producing rings 0--6 (Table~\ref{DistTab}).  {\it Right panel:} cumulative extinction from \citet{green19} in 8 equally spaced directions along the average position of ring 6 during Obs2 ($\sim$11.5$^{\prime}$ from \grb), extracted using the \texttt{dustmaps} software \citep{green18}.}
    \label{fig:3Dmaps}
\end{figure}

The \citet{lallement22} 3D map in the \grb\ direction displays 4 prominent extinction peaks between 400 and 750 pc (left panel of Fig.\,\ref{fig:3Dmaps}; \v Siljeg et al. in preparation) at distances compatible with those derived from the brightest rings detected in Obs1 (rings 1, 3, 4, and 6; top panel of Fig.\,\ref{PDs}). In particular, an extinction excess $\Delta$A$_V=0.4$ mag can be associated to the dust producing ring 6 (and possibly contributing also to ring 5),
whereas the contribution from the other dust layers is more difficult to disentangle. 
The angular resolution of the \citet{lallement22} map at such distance (730 pc) is
$\sim$2$^{\circ}$, but the spatial variability of the ring intensity on arcminute scale (Fig.\,\ref{fig:azimuth}) indicates that high resolution maps would be required. 
However, 
the \citet{green19} map, which has a better resolution,
cannot detect significant extinction peaks that can be safely associated to individual X-ray rings (right panel of Fig.\,\ref{fig:3Dmaps}).  

A significant extinction excess $\Delta$A$_V$=1.2 mag can instead be evaluated by selecting the voxels within 8$^{\prime}$ and 12$^{\prime}$ from \grb\ and at distances between 350 and 800 pc in the \citet{green19} map. The selected angular interval fully covers the 6 complete rings detected in Obs1 (rings 1--6) and the distance range includes all the dust layers producing them (Table~\ref{DistTab}). 
In the same distance interval, $\Delta$A$_V=2.1$ mag is derived from \citet{lallement22}.

Since the X-ray halo produced by dust closer than 350 pc was already 
outside the EPIC field of view during Obs1, the total amount of dust generating all the rings entirely observed by \xmm\ can be derived from the difference between the total Galactic extinction reported in 2D maps and the optical extinction measured up to 350 pc in 3D maps. The extinction from 2D maps, derived from optically thin dust emission, should be preferred to the integration of 3D maps over the full distance range, which is known to underestimate the extinction at large distances, where the properties of background stars are difficult to constrain. Assuming R$_V=3.1$, the average of the reddening values reported by \citet{schlafly11} between 2$^{\prime}$ and 12$^{\prime}$ from \grb, where all the \xmm\ rings were detected, translates into A$_V$= 3.9 mag. After the subtraction of the optical extinction within 350 pc, we estimate a residual extinction of 3.0 mag based on \citet{green19} and 3.5 mag according to \citet{lallement22}.

\begin{figure}
    \includegraphics[width=0.8\textwidth]{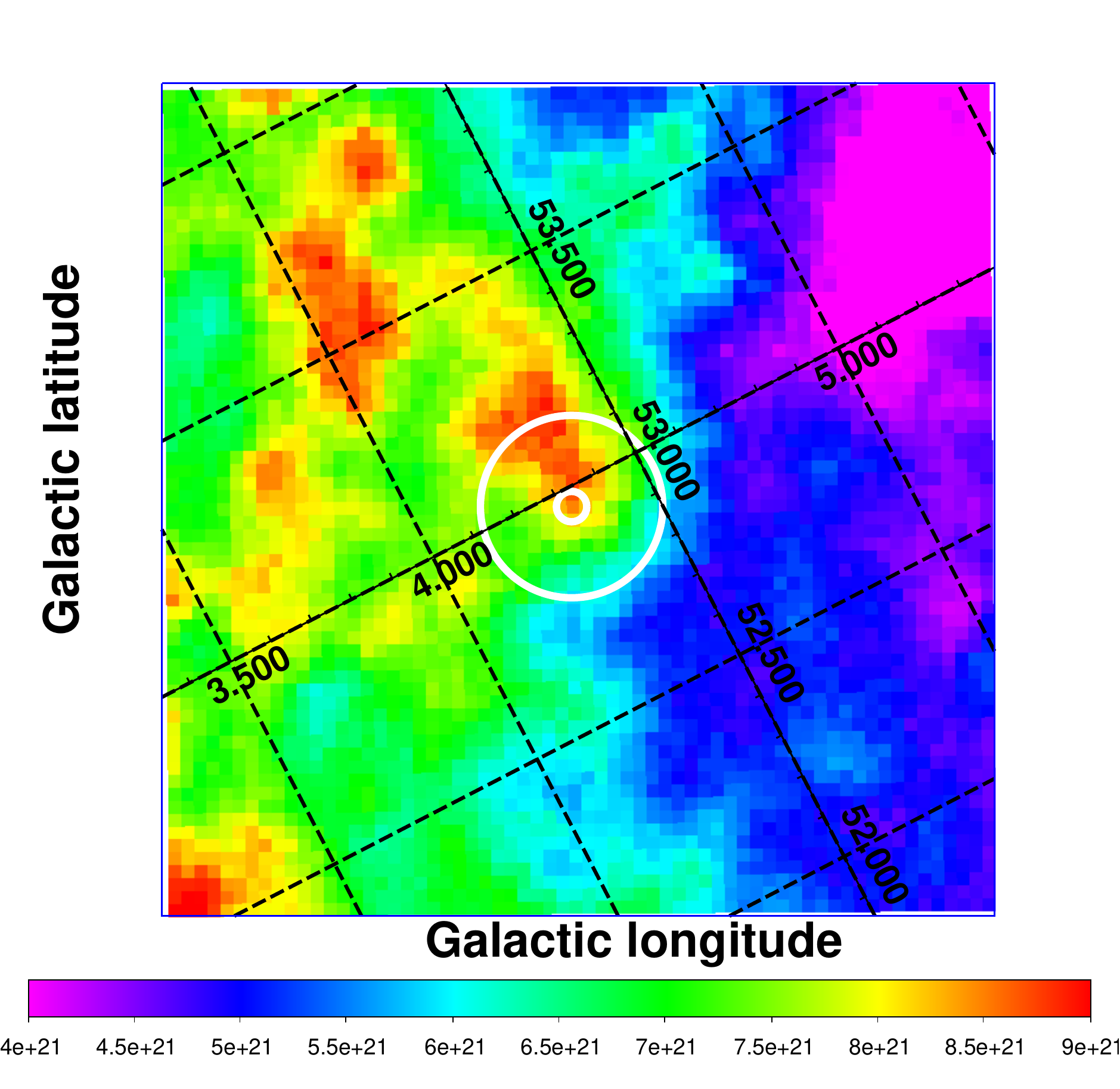}
\caption{Map of the total hydrogen column density in the sky area around \grb\ \citep{planck14}. The white circles (radii of 2$^{\prime}$ and 12$^{\prime}$) indicate the region covered by the X-ray rings during \xmm\ observations.} 
    \label{fig:planck}
\end{figure}

In the same 2$^{\prime}$--12$^{\prime}$ annulus, 
assuming a conversion factor of $8\times10^{25}$ cm$^{-2}$ from dust optical depth at 353 GHz to hydrogen column density, obtained from the analysis of thermal dust emission with the Planck satellite \citep{planck14}, we derived an average Galactic $N_{\rm H,G}$ toward GRB221009A of 
$7.4\times10^{21}$\,cm$^{-2}$ (see Fig.\,\ref{fig:planck}). This value, combined with the total optical extinction reported above, allows us to derive a conversion factor 
\begin{equation}
N_{\rm H}/A_V=1.9\times10^{21} {\rm cm}^{-2} {\rm mag}^{-1}. 
\label{NhAv}
\end{equation}

This factor is in the same range as the ratios obtained by averaging optical extinction and hydrogen column density derived from different samples of X-ray bright objects (e.g., \citealt{predehl95,watson11,zhu17}). This relation will be used in the spectral analysis to convert the extinction in the dust layers ($\Delta A_V$) into column density ($\Delta N_{\rm H}$).

\bibliography{sample631}{}
\bibliographystyle{aasjournal}



\end{document}